\documentclass[aps, prd, twocolumn, groupedaddress, showpacs, showkeys]{revtex4}
\usepackage{graphicx}
\usepackage{dcolumn}
\usepackage{mathrsfs}
\usepackage{amsmath}
\usepackage{latexsym}
\usepackage[usenames]{color}

\newcommand{\ben}{\begin{equation}}
\newcommand{\een}{\end{equation}}
\newcommand{\bea}{\begin{eqnarray}}
\newcommand{\eea}{\end{eqnarray}}
\newcommand{\ba}{\begin{array}}
\newcommand{\ea}{\end{array}}
\newcommand{\bit}{\begin{itemize}}
\newcommand{\eit}{\end{itemize}}

\def\math{\mathsurround 0pt}
\def\oversim#1#2{\lower.5pt\vbox{\baselineskip0pt \lineskip-.5pt
       \ialign{$\math#1\hfil##\hfil$\crcr#2\crcr{\scriptstyle\sim}\crcr}}}

\definecolor{Blue}{rgb}{0.0,0,1.0} 
\definecolor{Red}{rgb}{1.0,0,0.0} 

\begin{document}

\title{Radiation and Relaxation of Oscillons}
\author{Petja Salmi$^{1}$}
\email{petja.salmi@uct.ac.za}
\author{Mark Hindmarsh$^{2}$}
\email{m.b.hindmarsh@sussex.ac.uk}
\affiliation{$^{1}$Department of Mathematics and Applied Mathematics, 
University of Cape Town, Rondebosch 7701, Cape Town, South Africa \\
$^{2}$Department of Physics \& Astronomy, 
University of Sussex, Brighton BN1 9QH, UK}
\date{\today}

\begin{abstract}
We study oscillons, extremely long-lived localized 
oscillations of a scalar field, with three different potentials: 
quartic , sine-Gordon model and in a new class of convex potentials. 
We use an absorbing boundary at the end of the lattice to remove 
emitted radiation. 
The energy and the frequency of an oscillon evolve in time and 
are well fitted by a constant component and a decaying, radiative part 
obeying a power law as a function time. 
The power spectra of the emitted radiation show several 
distinct frequency peaks where oscillons release energy.
In two dimensions, and with suitable initial conditions, oscillons do not 
decay within the range of the simulations, 
which in quartic theory reach $10^8$ time units.
While it is known that oscillons 
in three-dimensional quartic theory and sine-Gordon model decay relatively 
quickly, we observe a surprising persistence of the oscillons in the convex 
potential with no sign of demise up to $10^7$ time units.
This leads us to speculate that an oscillon in such a potential 
could actually live infinitely long both in two and three dimensions.
\end{abstract}

\keywords{Solitons, Oscillons, Breathers}
\pacs{03.65.Pm, 11.10.-z, 11.27.+d}

\maketitle

\section{Introduction}

The theorem by Derrick states that there are no non-trivial static solutions 
of finite energy in scalar theories above one dimension~\cite{Derrick:1964ww}.
This does not say anything about time-dependent solutions. 
Example of such are Q-balls~\cite{Coleman:1985ki}, whose stability is 
guaranteed by a conservation law (for more recent studies 
on Q-balls see e.g.~\cite{Multamaki:1999an,Tsumagari:2008bv}) and that have 
been studied e.g. as a dark matter candidate 
(see for instance~\cite{Dine:2003ax}).
There is, however, no obvious reason why oscillations of a real scalar 
field can persist and stay localised, yet such configurations, 
{\it oscillons} exist in many models (we will use the term oscillon, 
presented originally in~\cite{Gleiser:1993pt}, in a slightly loose manner 
throughout this study).
There is a stable breather solution in one-dimensional sine-Gordon model 
that during its period deforms to a separated kink and anti-kink.
There is no such exactly stable solution in $\phi^4$ theory~\cite{Segur:1987mg}.
However, even though the solution radiates energy away, 
there remains a possibility that the radiation rate is so suppressed 
that it will take indefinitely long for the dissipation to complete.

Oscillons were first found already in 
the 70's~\cite{Bogolyubsky:1976nx,Bogolyubsky:1976sc} and then 
rediscovered in the 90's when the dynamics of phase transitions 
was studied~\cite{Gleiser:1993pt}. 
There has been substantial interest in oscillons recently. 
On the theoretical side of understanding the dynamics of oscillons, 
there have been a number of studies of oscillons of $\phi^4$ theory 
in small-amplitude 
approximation~\cite{Fodor:2006zs,Fodor:2008es,Fodor:2008du,Fodor:2009kf}. 
Furthermore, the attractor basin of oscillons and  its fractal nature 
have been studied both analytically~\cite{DiPlinio:2010dr} and 
numerically~\cite{Honda:2010gb}.
Very recently there has been interest in oscillons coupled to gravity, 
{\it oscillatons}~\cite{Fodor:2009kg,Grandclement:2011wz,ValdezAlvarado:2011dd}.
Oscillons have also been found in dilaton-scalar theories~\cite{Fodor:2009xw}.
A new class of solutions, called flat-top oscillons by the authors, 
were reported in~\cite{Amin:2010jq}.
While the body of work mentioned above has been on a classical level, 
the study in~\cite{Hertzberg:2010yz} considered quantum corrections 
oscillons are subject to, redeeming oscillons considerably less robust against 
the quantum effects than often previously has been assumed. 

Though stability or finite lifetime are of considerable theoretical interest, 
from the point of view of a many 
phenomenological consequences, it is not important if the non-trivial 
solutions are actually stable like the breather in one-dimensional 
sine-Gordon model as long as they are far longer-lived than the natural 
time scale related to the process on which they will have an effect. 
A natural realm of oscillons to appear is in the early Universe and 
such a process there could be baryogenesis. 
The role oscillons could potentially have on baryogenesis is providing 
the necessary non-equilibrium needed for successful creation of 
matter-antimatter asymmetry analogous to Q-balls 
whose cosmological impact has been studied 
extensively (for a review, see e.g.~\cite{Enqvist:2003gh}).
In this context it is very interesting that an oscillon solution 
has been found in 
the bosonic 
sector of the Standard Model ~\cite{Farhi:2005rz,Graham:2006vy,Graham:2007ds}. 
When the 
scalar and vector masses in the theory are set to 
be $m_{H} = 2m_{W}$, the oscillon is sufficiently long-lived 
that it has not been seen to decay in any numerical simulation to 
date~\cite{Farhi:2005rz,Graham:2006vy,Graham:2007ds}. 
Oscillons have also been found in two and three-dimensional 
Abelian-Higgs model in deep 
type I regime~\cite{Gleiser:2007te,Gleiser:2008dt}.
A number of studies have considered oscillons in the early Universe. 
While the oscillons in the Standard Model or Abelian-Higgs model are 
obviously objects in several 
fields, a recent study~\cite{Gleiser:2011xj} found that oscillons can 
also form in two scalar fields in models that are relevant from the point 
of view of hybrid inflation 
(on that topic see also~\cite{Broadhead:2005hn}).
Other recent studies of formation of oscillons after an inflationary epoch 
but in single field models include~\cite{Amin:2010dc,Amin:2011hj}.
Regarding the subsequent evolution of oscillons after their formation 
in a cosmological setting, both 
numerical~\cite{Graham:2006xs} and analytical~\cite{Farhi:2007wj} 
considerations in one dimension have shown that oscillons can persist 
for a considerably time in expanding backgrounds 
(see also~\cite{Amin:2010jq}). 
The proposal of oscillons facilitating vacuum tunneling in the context
of the string landscape~\cite{Saffin:2008vi} is an application quite far 
from the original idea of oscillons affecting the bubble nucleation process, 
the resonant nucleation studied e.g. in~\cite{Gleiser:2004iy} 
(for a study of bubble collisions, see also~\cite{Dymnikova:2000dy}). 
A recent study considered the formation of oscillons in the 
quintessence field at the present epoch of the Universe~\cite{Amin:2011hu}.

Lastly, regarding the significance of the oscillons, it should be 
mentioned that they are an interesting link between particles 
and solitons. Several numerical studies in one dimension 
have considered the formation of kink-antikink pairs. It has been found 
that scattering of breathers can create a kink and an 
antikink~\cite{Dutta:2008jt}. Oscillons can also decay to a kink and 
antikink when the potential of the model 
is distorted~\cite{Carvalho:2009ac}, as well as oscillons being an 
intermediate state in the process of formation of the 
kink-antikink pair~\cite{Romanczukiewicz:2010eg}.

In an earlier work~\cite{Hindmarsh:2006ur} we studied oscillons in 
two dimensions on the lattice with 
periodic boundary conditions.  
The aim of this study is to examine in detail the evolution of oscillons and 
the properties of the radiation they emit, in order to better understand the 
reasons for their longevity with the use of absorbing boundary conditions.  

Previous work in a theory with a quartic potential has identified stable 
periodic solutions with incoming radiation, called quasl-breathers, which 
are closely related to oscillons~\cite{Watkins,Saffin:2006yk}.  
There is a critical 
oscillation frequency 
at which the quasi-breather has the minimum 
energy~\cite{Watkins,Saffin:2006yk}, and once the incoming 
radiation is removed, the frequency of the resulting solution 
evolves towards its 
critical value according to a power law  
in time~\cite{Saffin:2006yk}. 
Power laws featured also strongly in~\cite{Gleiser:2008ty,Gleiser:2009ys} 
where the radiation rates of 
oscillons were studied, starting from the assumption of a strictly 
Gaussian form  for the oscillon, and a decay width calculated by comparison 
to linear theory. An equation relating the energy loss rate to the rate of 
change of the oscillon amplitude was derived, with power law solutions for 
the time evolution. 
Note that oscillons in the small-amplitude 
expansion~\cite{Fodor:2008du,Fodor:2009kf} behave differently, 
in that they shed energy in an exponentially suppressed way as the 
amplitude goes to zero, and do not obey a simple power law.

\begin{table}[tbp]
 \begin{tabular}{|c|c|}
     \hline
     model & potential $V(\phi)$ \\ \hline 
     quartic ($\phi^4$)  & $\displaystyle\frac{1}{4}(\phi^2 - 1)^2$ \\ 
     sine-Gordon & $\displaystyle\frac{1}{\pi^2}(1+\cos(\pi \phi))$ \\ 
     convex & $\displaystyle\frac{\phi^2}{1+\phi^{2p}}$ \\ \hline
   \end{tabular}
\caption{\label{t:models} Potentials for the three models considered in this 
paper. In the convex potential, $p$ is a real parameter, with $p <1$.  
For the data shown $p=0.45$.}
 \end{table}

In this paper we determine these power laws in three 
models in two and three dimensions, 
consisting of a single canonically normalised real scalar field $\phi$
with potentials listed in Table~\ref{t:models}, 
and examine the power spectra of the emitted radiation.  
Our power spectra show that for the quartic and sine-Gordon theories 
the radiation 
is predominantly emitted at an integer multiple of the basic oscillation 
frequency, either 3 or 2 depending 
whether the potential is symmetric or not.
For the convex potential, the radiation is emitted just above the 
threshold set by the mass parameter.
A fundamental assumption 
made in~\cite{Gleiser:2009ys} in 
developing a theory of oscillons in 
the quartic theory was that the radiation 
is predominantly emitted just above threshold, which we see is incorrect.  
Nonetheless, good fits to the time 
evolution of the energy, frequency and amplitude were obtained in that work.

An important technical development in our work is the construction 
of absorbing boundary conditions for a massive field, 
which provides an economical alternative to the adiabatic damping 
technique, introduced in~\cite{Gleiser:1999tj}, and 
also used in \cite{Saffin:2006yk}, which sets a location dependent 
friction co-efficient on the lattice.
Absorbing boundaries provide an alternative 
approach to remove the dispersive waves from the lattice, 
requiring no extra lattice sites for its operation.

The paper is organised as follows.
The numerical set-up is briefly reviewed in the following section.
While the equations of the absorbing boundary conditions are 
presented in the appendix we briefly discuss their implications for 
simulations together with the data obtained using 
quartic theory as an example.
We then proceed to present the results  for the power laws and 
the radiation power spectra 
in the three theories in two and three dimensions. 
We find a surprising stability for the oscillon in the convex potential, 
which leads us to conjecture that it may be stable in three dimensions.

\section{Numerical Set-up}

\label{s:Models}

The Lagrangian for a single real scalar field~$\phi$ 
is given by
\begin{eqnarray}
 \mathscr{L} =
\frac{1}{2} {\partial}_{\mu} \phi {\partial}^{\mu} \phi - V(\phi), 
\label{lagrangian}
\end{eqnarray}
and the equation of motion thus reads
\begin{eqnarray}
\ddot{\phi}-\nabla^{2}\phi + V'(\phi)=0. 
\label{eqm}
\end{eqnarray}
This field equation is evolved using a leapfrog update and 
a three-point spatial Laplacian accurate to O($dx^2$).
We perform (1+1)-dimensional simulations assuming radial 
symmetry. 
These permit the choice of a much finer lattice spacing compared 
with (2+1)-dimensional simulations carried out in a previous 
study~\cite{Hindmarsh:2006ur}. We kept the ratio of time step and 
lattice spacing fixed and matching to 
the one used in (2+1)-dimensional simulations, $dt:dx = 1:5$. 
The size of the one-dimensional grid was chosen so 
that the physical distance from the core of the oscillon to the absorbing 
boundary was always $100$ units. 
The simulations are performed on a grid of 10,001 lattice 
points and $dx=0.01$, $dt=0.002$.

We set the field $\phi$ and the field momentum $\Pi = \dot{\phi}$ at the 
origin $r=0$ to have the same value as on the next lattice site with
a non-zero value of the radius $r$. 
The updates used at the far end of the grid, where 
the absorbing boundary is located, are derived in the appendix, while we 
comment the simulations with the absorbing boundaries shortly 
in the next section.
We use three consecutive points in the grid to evaluate the gradients with 
the precision also accurate to O($dx^2$).


\section{Quartic Potential}

We study here the degenerate double-well quartic potential 
which defines the model that we refer also as $\phi^4$ theory
\begin{eqnarray}
V(\phi)=\frac{1}{4} \lambda (\phi^2-v^2)^2 \, ,
\label{quartic} 
\end{eqnarray}
which sets the mass of the theory to be $m^2 = 2 \lambda v^2$. 
Scaling out the vacuum expectation value and coupling the potential 
can be written
\begin{eqnarray}
V(\phi)=\frac{1}{4} (\phi^2 - 1 )^2 
\label{scaledquartic}
\end{eqnarray}
so that the minima are located at $\phi = \pm 1$ and 
the local maximum at $\phi = 0$.
We use the following Gaussian initial data to create an oscillon 
\begin{eqnarray}
\phi(r) =  \,\left( 1 - C \cdot \exp(-r^2/r_{0}^2)\right),
\label{gaussian-ansatz}
\end{eqnarray}
where $r$ is the distance to the center of an oscillon. 
The width of the distribution is set to be 
$r_0 \simeq 2.9$ (in units of  $(\sqrt{\lambda}\, \eta )^{-1}$) in two and 
$r_0 \simeq 3.0$ in three dimensions, 
suggested optimal choices in~\cite{Copeland:1995fq}. The 
maximum displacement in two dimensions is set to be $C=1$ so that the center of 
the oscillon starts from the local maximum of the potential, while we 
use $C=-1$ in three dimensions again following~\cite{Copeland:1995fq}.

\subsection{Properties in $\phi^4$ theory in two dimensions}

\begin{figure}

\centering

\includegraphics[width=0.94\hsize]{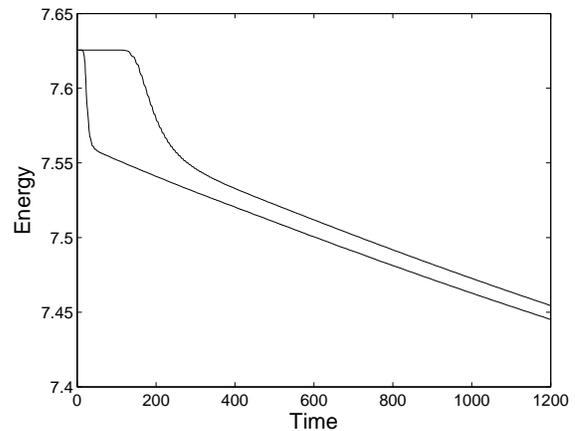}
 \caption{\label{f:quartic_spherical_energy0} The total energy in 
the lattice (top) and energy inside a shell of radius $r=5 r_{0}$ 
(bottom) in $\phi^4$~theory in two dimensions.}
\end{figure}

As we kick off the oscillon by the somewhat arbitrary 
initial condition~(\ref{gaussian-ansatz}), it naturally sheds at least some 
fraction of the energy in the form of propagating 
waves moving away. As long as that emission is not too violent and at least 
further distance away from the oscillon core spreading and consequently damped 
in two or higher dimensions 
these waves can be assumed to be well-characterized just by a linear 
approximation of the equation of motion~(\ref{eqm}), see~(\ref{appendix1}). 
Once the wave reaches the boundary of the grid it is absorbed and the energy 
carried by the wave removed from the lattice. We thus expect monotonically 
decreasing total energy in the system, 
and this is also what we observe. Figure~\ref{f:quartic_spherical_energy0} 
shows the total energy (top line) in the lattice at the very early stage 
of the simulation.
The total energy in the lattice will always include the radiative 
component emitted by the oscillon which has not yet reached the boundary 
of the grid. 
Thus we measure also the energy inside the radius $r=5r_{0}$ from the center 
of the oscillon and refer this as the energy of the oscillon. The oscillon is 
generally well localised inside this volume, but it covers less 
than $15\%$ of the lattice used here. The energy inside this shell is 
also depicted in Figure~\ref{f:quartic_spherical_energy0} (bottom line).
In the beginning it takes a finite time the wave to reach the boundary 
and even then the 
decrease in total energy is more gradual compared with the energy inside 
the shell; this is caused by the dispersion of the emitted waves. However, 
after the initial transient phase these two ways to measure energy track each 
other well; there is only a time delay between them.
This also provides a crucial check for the adequacy of the absorbing boundary 
conditions we use. Any reflected waves will travel back and reach the inner 
shell consequently increasing energy measured inside the radius there. 
There is no sign of such a burst at a visible level 
in Figure~\ref{f:quartic_spherical_energy0} and we expect only a 
very minor reflection taking place at the boundary. 
We monitor the total energy and the 
energy inside the aforementioned shell throughout the simulations.

\begin{figure}

\centering

\includegraphics[width=0.94\hsize]{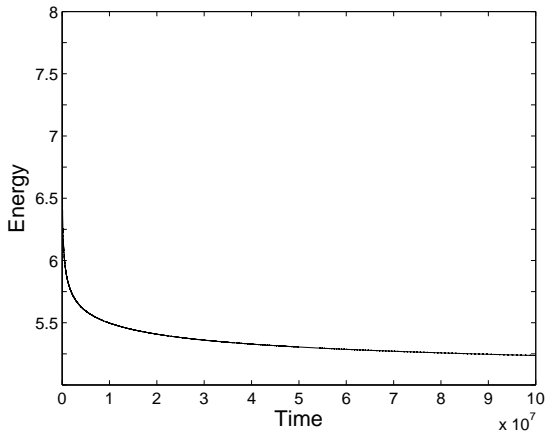}
 \caption{\label{f:quartic_spherical_energy1} The energy inside a 
shell of radius $r=5 r_{0}$ in $\phi^4$~theory in two dimensions over 
the span of the simulation up to $10^8$~time units.}
\end{figure}

\begin{figure}

\centering

\includegraphics[width=0.94\hsize]{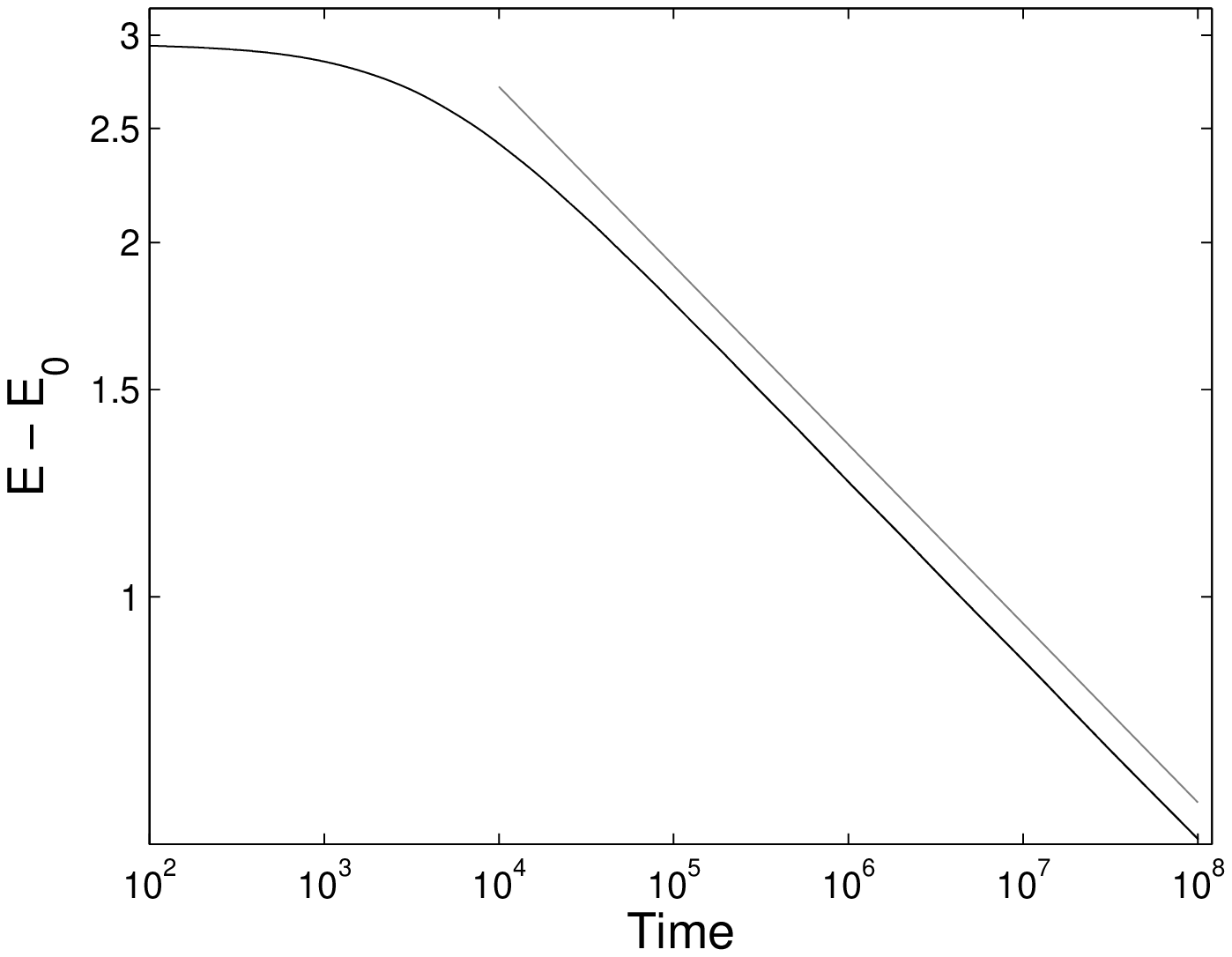}
 \caption{\label{f:quartic_spherical_energy} The difference of the 
energy inside a shell of radius $r=5 r_{0}$ and the 
constant~$E_{0} = 4.613$ in $\phi^4$~theory in two dimensions. 
The grey straight line is a guide to eye of a power 
law $(E - E_{0}) \sim t^{-\delta}$ with the slope $\delta=0.152$.}
\end{figure}

While at the early stage there is seemingly linear decrease in energy of 
the oscillon as shown in Figure~\ref{f:quartic_spherical_energy0}, this rate of 
energy loss flattens out.
Figure~\ref{f:quartic_spherical_energy1} shows the energy inside the 
radius $r=5r_{0}$ over the simulation that reaches over $10^8$ time units. 
First of all, there is no sign of the demise of the oscillon by this 
point in time.
The energy is naturally monotonically decreasing function of time, 
but the rate of decrease slows down raising the question if the energy 
approaches asymptotically a constant value, which we denote by $E_0$ from 
now on. We searched for a power law of the 
form $\left(E(t) - E_0\right) \sim t^{-\delta}$.
In practice we performed least-square fit with three free parameters, 
the asymptotic energy~$E_0$, the exponent~$\delta$ and an amplitude in 
the power law for the data.  
The best fit for the data presented in 
Figure~\ref{f:quartic_spherical_energy1} is 
shown in Figure~\ref{f:quartic_spherical_energy} where we plot the 
difference of the energy inside the shell and the constant $E_0 = 4.613$ 
on a logarithmic scale. The grey line is a guide to eye of a slope $-0.152$. 
There is very good agreement to this over three orders of magnitude 
in time providing strong evidence for the power law decay of the radiative 
component governed by a small exponent 
$\delta = 0.1520 \pm 0.0005$ (we quote hereafter uncertainties in 
quantities based on $68 \%$ confidence intervals derived from standard 
regression analysis applied to the subset of the data used for the fit).  
Moreover, there exist asymptotic value for the energy $E_0$ of the 
oscillon with a value $E_0 = 4.613 \pm 0.003$, 
thus $E_0 \gtrsim 4.6$.

\begin{figure}

\centering

\includegraphics[width=0.94\hsize]{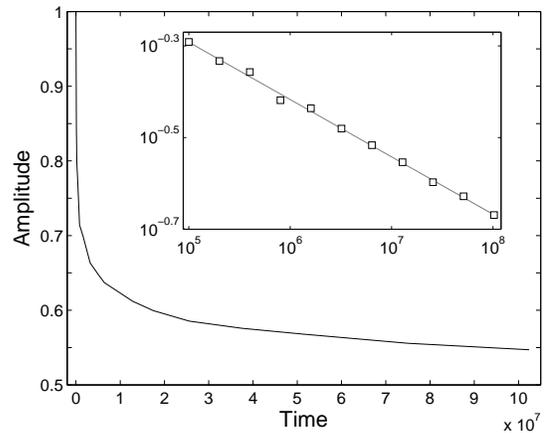}
 \caption{\label{f:quartic_amplitude} The amplitude as a 
function of time in $\phi^4$~theory in two dimensions. The inset 
shows the data points in the fit $A - A_0$ for 
$A_0=0.33$ for time greater than $10^5$. The grey straight 
line is the best fit power law 
$(A-A_{0}) \sim t^{-\varrho}$ with 
the slope $\varrho=0.13$.}
\end{figure}

\begin{figure}

\centering

\includegraphics[width=0.94\hsize]{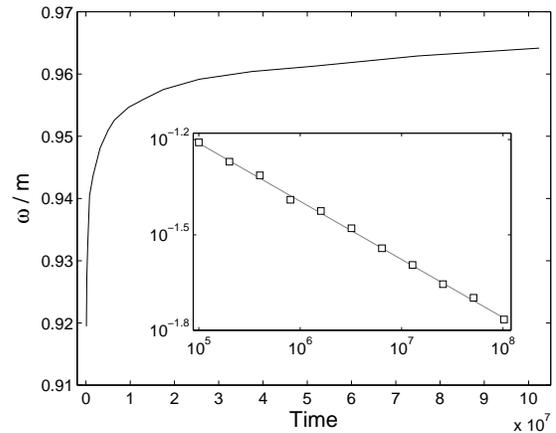}
 \caption{\label{f:quartic_frequency} The oscillation frequency as a 
function of time in $\phi^4$~theory in two dimensions. The inset 
shows the data points in the fit $\omega^{\star}-\omega$ for 
$\omega^{\star}=0.981$ for time greater than $10^5$. The grey straight 
line is the best fit power law 
$(\omega^{\star} - \omega_{0}) \sim t^{-\gamma}$ with 
the slope $\gamma=0.182$.}
\end{figure}

The decrease in energy is a consequence of the decline in the 
amplitude of the oscillations, i.e. the maximum excursion the field 
makes at the center of the oscillon, which we take to be towards the 
local maximum of the potential, initially set by~$C=1$. 
The amplitude $A$ as a function of time 
is shown in Figure~\ref{f:quartic_amplitude}. 
Also this quantity is fitted relatively well by a power 
law, $\left(A(t) - A_0\right) \sim t^{-\varrho}$. The best-fit 
values for the asymptotic value of the amplitude is $A_0= 0.33 \pm 0.06$ 
and for the exponent governing the power law $\varrho = 0.13\pm 0.02$. 
The comparison between the best-fit and the data is shown in the inset in 
Figure~\ref{f:quartic_amplitude}. This suggests 
that the oscillations would still undergo a substantial decrease 
in size before reaching the asymptotic value $A_0$. It should be 
noted that as the amplitude is not an integrated quantity like the 
energy of the oscillon, the uncertainty in the fit is prominently larger, 
and there remains scatter as can be seen in the inset 
in Figure~\ref{f:quartic_amplitude}.

With decreasing amplitude the frequency of the oscillon increases. 
We measure the frequency through the oscillation period, determined 
from the field's three consecutive crossings of the minimum of the potential 
at the center of the oscillon. Based on the time step used we expect 
the accuracy in determining the frequency from the data to be around $0.001$. 
Figure~\ref{f:quartic_frequency} shows the frequency as a function of time 
over the simulation. At the end of the simulation the 
frequency is $\omega = 0.964$, still considerably below the threshold 
for radiation, $\omega/m=1$.

Previous numerical studies of quasi- or pseudo-breathers have established 
that there exists a critical frequency~$\omega^{\star}$ such that 
$\omega^{\star} < 1$ and above which the oscillon disintegrates 
when the dimensionality in $\phi^4$ theory is more than 
two~\cite{Watkins,Saffin:2006yk}. Furthermore, 
it has been suggested that the approach of the frequency towards its 
critical value is governed by a power law as 
$\left(\omega^{\star} - \omega(t)\right) \sim t^{-\gamma}$ 
in~\cite{Saffin:2006yk}. 
Thus we searched again a power law by performing the 
least-square fit to the data. 
The best fit yields the critical 
frequency $\omega^{\star} = 0.981 \pm 0.003$ and 
the exponent governing the power law $\gamma = 0.18 \pm 0.02$.
The data points for $\omega^{\star} - \omega(t)$ are 
depicted in the inset of Figure~\ref{f:quartic_frequency}.
There remains a considerable scatter from the central value, the grey 
straight line that demonstrates the best-fit power law~$t^{-\gamma}$, 
but there is evidence for a critical frequency $\omega^{\star} < 1$.

\begin{figure}

\centering

\includegraphics[width=0.94\hsize]{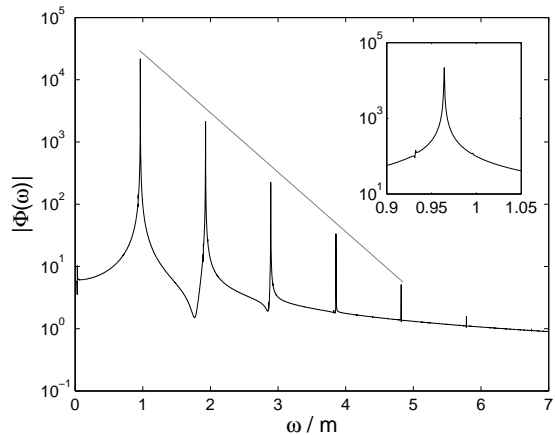}
 \caption{\label{f:quartic_spectra} The power spectrum of the field at 
the center of the oscillon, $r=0$, in $\phi^4$~theory in two dimensions. 
The grey straight line is a guide to eye of the exponential $\exp(-b\,\omega)$ 
with the slope $b=2.2$. The inset shows the spectrum around the oscillation 
frequency $\omega_0 \simeq 0.96$.}
\end{figure}

\begin{figure}

\centering

\includegraphics[width=0.94\hsize]{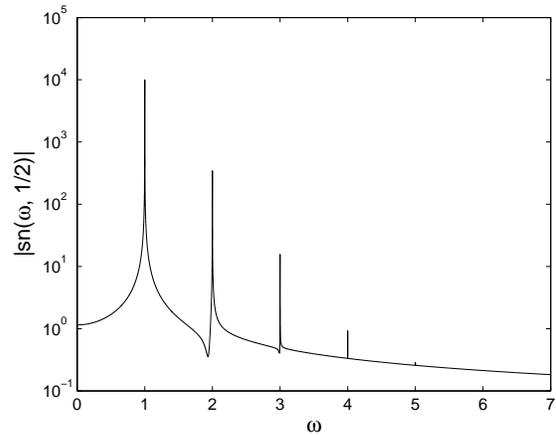}
 \caption{\label{f:jacobi} The Fourier transform of the 
Jacobi function ${\rm dn}(t,1/2)$. The height of the first peak is 
normalised to $10^4$ and its location to unity.}
\end{figure}

\begin{figure}

\centering

\includegraphics[width=0.94\hsize]{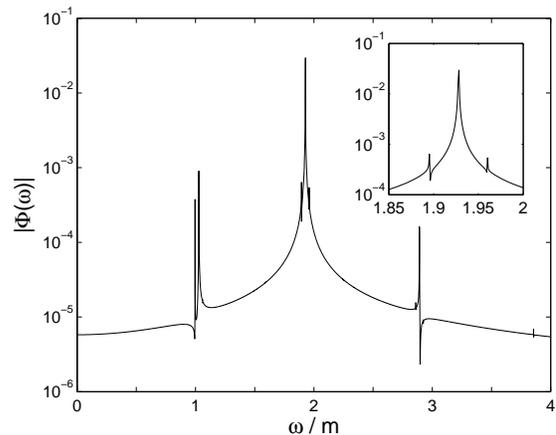}
 \caption{\label{f:quartic_spectra2} The power spectrum of the field 
far away from the oscillon core at $r=80$, in $\phi^4$~theory in two 
dimensions. The inset shows the spectrum around the location of the most 
prominent peak at $\omega \simeq 1.93$.}
\end{figure}


In addition to the oscillation frequency, the whole power spectrum 
of an oscillon is of interest and insight to the properties of 
oscillons can be gained by a study in frequency space.
We denote the oscillation frequency 
now by $\omega_{0}$ to distinguish it from other frequencies present 
and secondly, to highlight that it can be considered a constant for 
any practical purposes in relatively short time intervals considered here. 
A separable ansatz consisting of a sum of 
multiplies of this frequency each with a separate spatial dependence 
has seen to converge 
quickly~\cite{Watkins,Honda:2001xg,Saffin:2006yk}
\begin{eqnarray}
\phi(r,t) = \sum_{n=0}^{\infty} f_{n}(r) \cos(n \omega_{0} t) \,.
\label{solution-ansatz}
\end{eqnarray}

In an earlier work~\cite{Hindmarsh:2007jb} we used a method inspired 
by one-particle spectral function at zero momentum in classical 
approximation (for spectral function see~\cite{Aarts:1997kp,Aarts:2001yx}). 
This technique, employed to trace oscillons moving with varying speed, 
could be utilised here as well. As the spectral function involves a volume 
average it produces generally quite a clean signal and reduces background 
noise. However, we wish to study the frequencies present at different 
locations on the grid and thus choose to obtain the straightforward 
Fourier transform of the field~$\phi(r)$ on a fixed lattice site over 
a time interval.

Figure~\ref{f:quartic_spectra} shows the Fourier transform of the 
field at the center of the oscillon, $\phi(r=0)$ over the interval of 
length 5000 in time units starting at time~$t=1.024 \cdot 10^8$. 
The oscillation frequency $\omega_{0}$ (thus at this time $\omega_0=0.964$) 
is naturally the most pronounced mode, its multiples appearing in the 
spectra with starkly smaller amplitudes. 
This suppression is clearly exponential, illustrated by the grey straight 
line superimposed above the peaks in Figure~\ref{f:quartic_spectra}. 
One should note here that when performing the discrete Fourier transform in 
a limited time interval it is likely that we do not capture the height 
of the very narrow peaks very precisely. The exponential fit to the data 
with the slope $-2.2$ should be thus considered rather indicative than 
quantitative. The results are qualitatively very similar to those obtained 
in~\cite{Hindmarsh:2006ur} where periodic boundary conditions were utilised 
in a two-dimensional lattice.
However, the signal here is very clean; there is hardly any 
noise present, but the curve shown is a smooth line. 
We point out that fast Fourier transform of Jacobi functions in an 
interval yields a very similar signal, including the exponential suppression 
of the amplitudes at which the higher multiples of the basic 
frequency appear as well as similar patterns between the peaks. 
An example is given in Figure~\ref{f:jacobi} where 
Jacobi function ${\rm dn} \left(\omega,1/2\right)$ is shown.
The resemblance is not co-incidental - Jacobi functions are solutions to 
a type of elliptic non-linear differential equations that result also 
by omitting the 
spatial dependence in the equation of motion~(\ref{eqm}) in $\phi^4$ theory. 
Elliptic partial differential equations have also been shown to 
describe quasi-breathers in $\phi^4$ theory~\cite{Fodor:2008es}.

The use of absorbing boundary on the grid offers the oppourtunity to 
examine the outgoing radiation far away from the oscillon and circumvent the 
influence of the frequency directly related to oscillations 
at~$\omega=\omega_0$. 
Figure~\ref{f:quartic_spectra2} shows the Fourier transform of the field 
in the same time 
interval as in Figure~\ref{f:quartic_spectra}, but at location $r=80$. 
This choice guarantees that the observation point is a considerable 
distance away from the oscillon core, but not at the edge of the 
lattice either where the equation of motion~(\ref{eqm}) 
is modified to create the absorbing boundary. 
There is signal at radiation frequency $\omega=1$ and just above it, but 
the most prominent peak is at the frequency $\omega \simeq 1.93$, 
thus coinciding with the second frequency $2\omega_0$ in the 
expansion~(\ref{solution-ansatz}). There is 
a peak in the spectra also at $3\omega_0$ and just a visible structure at 
$4\omega_0$, but the signal at $\omega=2\omega_0$ is the most dominant, 
the height of the peak being over thirty times larger than that at other 
frequencies, though still almost six orders of magnitude below the main 
peak at the oscillon core. 
This observation is evidence for that it is the first mode 
in~(\ref{solution-ansatz}) for which the frequency is
above the radiation threshold (here $n=2$) that is primarily responsible for 
carrying away the energy from the oscillon.

\subsection{Properties in $\phi^4$ theory in three dimensions}

\begin{figure}

\centering

\includegraphics[width=0.94\hsize]{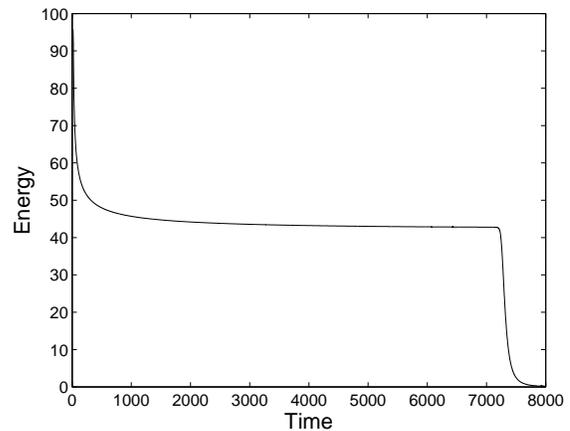}
 \caption{\label{f:quartic_spherical_energy0_3d} The energy inside a 
shell of radius $r=5 r_{0}$ in $\phi^4$~theory in three dimensions.}
\end{figure}

\begin{figure}

\centering

\includegraphics[width=0.94\hsize]{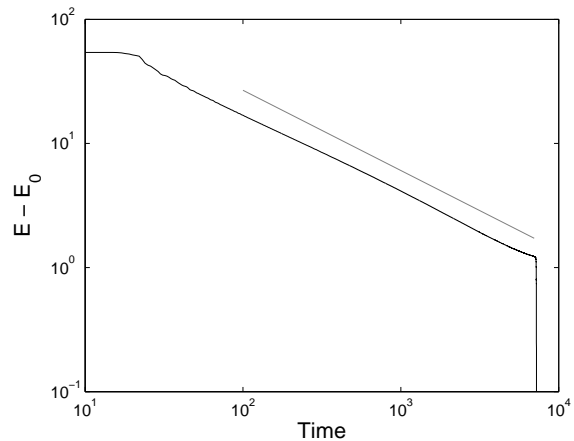}
 \caption{\label{f:quartic_spherical_energy_3d} The difference of the energy 
inside a shell of radius $r=5 r_{0}$ and the 
constant~$41.56$ in $\phi^4$~theory in three dimensions. 
The grey straight line is a guide to eye of a power 
law given by $(E - E_{0}) \sim t^{-\delta}$ 
with the slope $\delta=0.65$.}
\end{figure}

In three dimensions we start with a slightly broader width but 
considerably more energetic initial profile than in two dimensions 
as we set $C=-1$ and thus the amplitude 
of the excursion away from the minimum is towards the steeper side of 
the potential. The energy inside the radius $r=5r_0$ throughout the 
simulation is shown in Figure~\ref{f:quartic_spherical_energy0_3d}.
There is a very abrupt drop in energy at the very beginning, 
but then the decrease slows down 
and energy stabilises above the value $E \approx 40$ for a considerable period 
of time, until the oscillon disintegrates around time $t=7200$ which appears 
in the data as the second stark drop in energy. 

Even though the oscillon clearly has  a finite lifetime here, the 
steady decrease in energy is relatively well captured by a power 
law. This is demonstrated in Figure~\ref{f:quartic_spherical_energy_3d} 
where the difference between the energy inside the shell and a 
constant~$E_0=41.56$ is shown on a logarithmic scale. The best-fit value 
for the exponent is~$\delta= 0.65 \pm 0.01$, thus the strength of the 
emitted radiation decays far faster in time than in two dimensions. 
This is naturally only of secondary importance because the asymptotic 
value for the energy $E_0 = 41.56 \pm 0.05$ is well below the value when 
the oscillon destabilises, which is approximately at $E=42.7$.

\begin{figure}

\centering

\includegraphics[width=0.94\hsize]{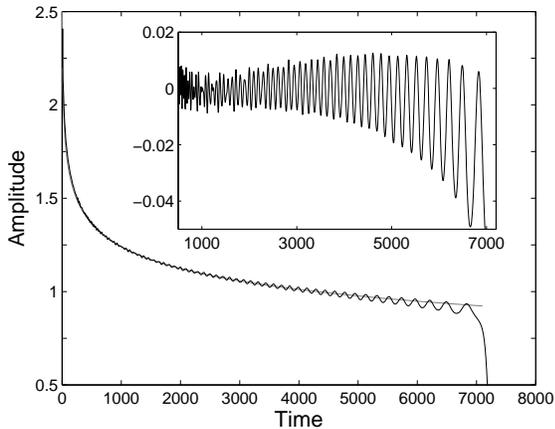}
 \caption{\label{f:quartic_amplitude_3d} The amplitude as a function of 
time in $\phi^4$~theory in three dimensions. 
The grey line is a guide to eye of a power law given 
by $(A - A_{0}) \sim t^{-\varrho}$ 
with $A_0 = -1.374$ and $\varrho=0.066$. 
The inset shows the residuals in the afore mentioned fit.}
\end{figure}

\begin{figure}

\centering

\includegraphics[width=0.94\hsize]{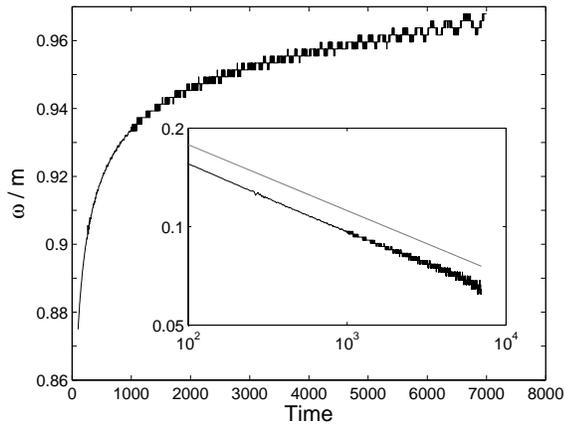}
 \caption{\label{f:quartic_3d_frequency} The oscillation frequency 
as a function of time in $\phi^4$~theory in three dimensions. 
The inset shows the data points in the fit $\omega^{\star} - \omega$ for 
$\omega^{\star} = 1.030$ for time greater than $100$.
The grey straight line is a guide to eye of a power law 
$(\omega^{\star} - \omega) \sim t^{-\gamma}$ 
with the slope $\gamma=0.2$.}
\end{figure}

The evolution of the amplitude $A$ of the oscillations is shown 
in Figure~\ref{f:quartic_amplitude_3d}. 
On top of the general downward trend that follows closely the time 
evolution of the energy, the most notable feature is the growing 
fluctuation in this quantity in the course of time. This amplified 
oscillation in the amplitude around the central value with a decreasing 
frequency makes a simple power law fit to the data challenging.
The best-fit to the data in the interval $500 < {\rm time} < 5000$ 
yields $A_0 = -1.4 \pm 0.3$ and $\varrho = 0.066 \pm 0.007 $, but 
it should be emphasized that the values in the fit are very sensitive 
to the selection of the time interval. 
The residuals of the fit are shown on a linear scale in the inset 
in Figure~\ref{f:quartic_amplitude_3d}. 
This highlights further the growing fluctuations, the beat, in the 
amplitude; to capture the time evolution of the amplitude 
entirely and adequately a model with extra parameters to account for 
these oscillations would be required.
The negative value of the 
asymptotic amplitude $A_0$ indicates strongly the finite lifetime of the 
oscillon.

\begin{figure}

\centering

\includegraphics[width=0.94\hsize]{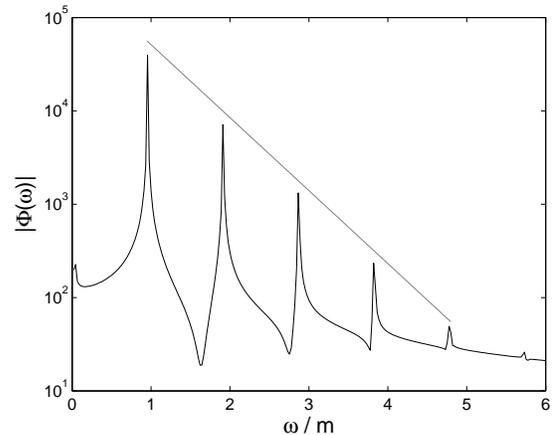}
 \caption{\label{f:quartic_3d_spectra} The power spectrum of the field at 
the center of the oscillon, $r=0$, in $\phi^4$~theory in three dimensions. 
The grey straight line is a guide to eye of the exponential $\exp(-b\,\omega)$ 
with the slope $b=1.8$.}
\end{figure}

\begin{figure}

\centering

\includegraphics[width=0.94\hsize]{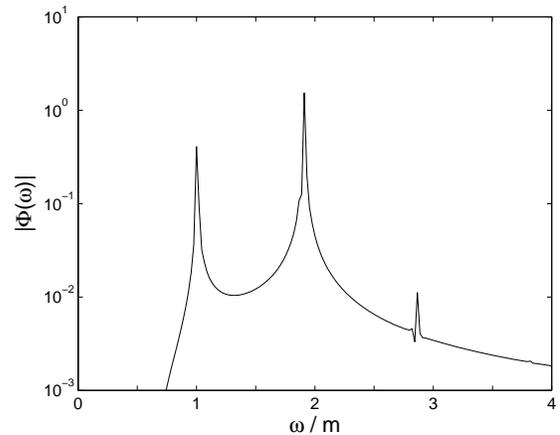}
 \caption{\label{f:quartic_3d_spectra2}  The power spectrum of the field 
far away from the oscillon core at $r=80$, in $\phi^4$~theory in three 
dimensions.}
\end{figure}

The oscillation frequency as a function of time is shown 
in Figure~\ref{f:quartic_3d_frequency}. 
The frequency approaches $\omega \simeq 0.97$ towards the end of the 
simulation before the oscillon decays. 
The fluctuations in the amplitude 
are naturally imprinted into the frequency. However, these do not prevent 
reaching a reasonable power 
law fit to the data over the whole span of the oscillon life time. 
In the best-fit the asymptotic value of the frequency is 
$\omega^{\star} = 1.030 \pm 0.003$ and the exponent governing the power 
law $\gamma = 0.20 \pm 0.01$. The power law is demonstrated in the 
inset in Figure~\ref{f:quartic_3d_frequency}. Like in the case of the 
amplitude, the asymptotic value of the frequency $\omega^{\star}$ 
remains physically meaningless as $\omega^{\star}>1$; the oscillon has 
a finite lifetime and decays even before the frequency reaches the 
threshold for radiation $\omega = 1$. 
For more extensive studies of oscillons in quartic theory in three 
dimensions, in particular the effect of the initial radius and amplitude, 
see~\cite{Gleiser:2009ys}.

While the decay of the oscillon in three dimensions restricts the length 
of the time interval we can observe the evolution, this fate merely 
enhances the importance to examine the frequency spectrum of the 
oscillon before its demise.
Due to the relatively fast energy loss the oscillation frequency 
shifts steadily and to overcome this effect we perform the Fourier 
transform over the interval of length $200$ in time units 
(the choice of the shorter interval naturally compromises the 
resolution we can expect on the frequency axis compared with the 
analysis in two dimensions).
We choose to start the interval at time $t=4000$, thus this interval 
is in the middle 
of the phase of relative subdued radiation appearing as the plateau in 
energy in Figure~\ref{f:quartic_spherical_energy0_3d}. 

The power spectrum at the center of the 
oscillon, $r=0$, is shown in Figure~\ref{f:quartic_3d_spectra}. 
It is qualitatively similar to the spectrum in two 
dimensions (Figure~\ref{f:quartic_spectra}) with peaks located at 
multiples of the oscillation frequency, which is here located approximately 
at $\omega_0 \simeq 0.955$.
There are even similar patterns between the peaks as those seen 
in two dimensions. 
In addition, the suppression of the amplitudes which the peaks appear 
is still exponential as well, the best fit value of the slope 
being $-1.8$, very comparable with the value obtained in the 
two-dimensional theory.
Unsurprisingly, the form of the power spectrum is dictated by the potential 
rather than the dimensionality of the theory.

The major difference between two and three dimensions is the far 
larger emitted radiation taking place in the latter resulting to a greatly 
larger width of the peaks in frequency space. This has been considered 
analytically in~\cite{Gleiser:2008ty,Gleiser:2009ys} in the case of the 
oscillation frequency. 

Figure~\ref{f:quartic_3d_spectra2} shows the Fourier transform of the field 
$\phi$ obtained in the 
same time interval as in Figure~\ref{f:quartic_3d_spectra}, but at the 
location $r=80$. 
The strength of radiation shows simply in the far larger absolute values of
the amplitudes compared with the situation in two dimensions, 
see Figure~\ref{f:quartic_spectra2}. 
There are three distinctive peaks present in the spectrum. 
There is a substantial structure rising immediately above the threshold for 
radiation at $\omega=1$. The origin of it can be traced to the 
large width of the peak of the basic oscillation frequency in 
Figure~\ref{f:quartic_3d_spectra}, 
the part of this peak for which $\omega>1$ overflows in the form of 
free radiation (for a quantitative study, see~\cite{Gleiser:2009ys}).
The radiation signal at $\omega \simeq 1$ is far stronger than in two 
dimensions. 
However, also in three dimensions the dominant peak is at the frequency 
corresponding the double of the oscillation frequency, $2\omega_0$.
The height of this peak is several factors larger than the one 
at $\omega=1$. Moreover, as waves moving at 
frequency $\omega=2\omega_0$ carry far larger amount of energy than 
those barely in the free, propagating domain at $\omega \gtrsim 1$, we 
conclude that the dominant channel for the energy loss in three 
dimensions is also through the frequency band at $\omega \simeq 2\omega_0$.
Finally, there is inferior, but still a visible peak in the spectra 
at $\omega = 3\omega_0$.


\section{sine-Gordon model}

Sine-Gordon model is of considerable interest from the point of 
view of oscillating solutions both on theoretical and phenomenological 
grounds: in one-dimension it has a stable breather solution and it is 
also the potential of axion field 
(formation of oscillating energy concentrations in axion field were 
studied in~\cite{Kolb:1993hw}).
In~\cite{Piette:1997hf} a fine-tuned initial ansatz was reported to create 
almost dissipationless breather in two dimensions.
Here we start with the same Gaussian ansatz as in $\phi^4$ theory given 
by~(\ref{gaussian-ansatz}) with $r_0\simeq 2.9$ and $C=1$. 
We scale the potential of sine-Gordon model 
\begin{eqnarray}
V(\phi)= \frac{\Lambda^4}{\alpha^2} \big( 1 - \cos (\alpha \phi) \big)
 \label{sine-Gordon}
\end{eqnarray}
to
\begin{eqnarray}
V(\phi)=  \frac{1}{\pi^2}\big( 1 + \cos (\pi \phi) \big) \, ,
 \label{scaled-sine-Gordon}
\end{eqnarray}
so that locations of the minima are the same as in $\phi^4$ theory, but 
here the mass $m^2=1$.

\subsection{Properties in two dimensions}

\begin{figure}
\begin{center}

\includegraphics[width=0.94\hsize]{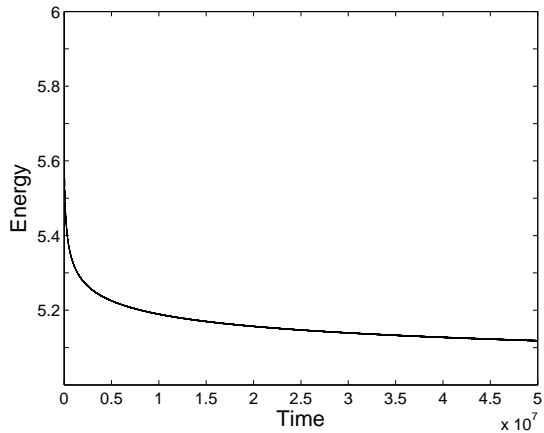}
 \caption{\label{f:sG_spherical_energy0} The energy inside a 
shell of radius $r=5 r_{0}$ in sine-Gordon model in two dimensions over 
the span of the simulation up to $5 \cdot 10^7$~time units.} 

\end{center}
\end{figure}

\begin{figure}
\begin{center}

\includegraphics[width=0.94\hsize]{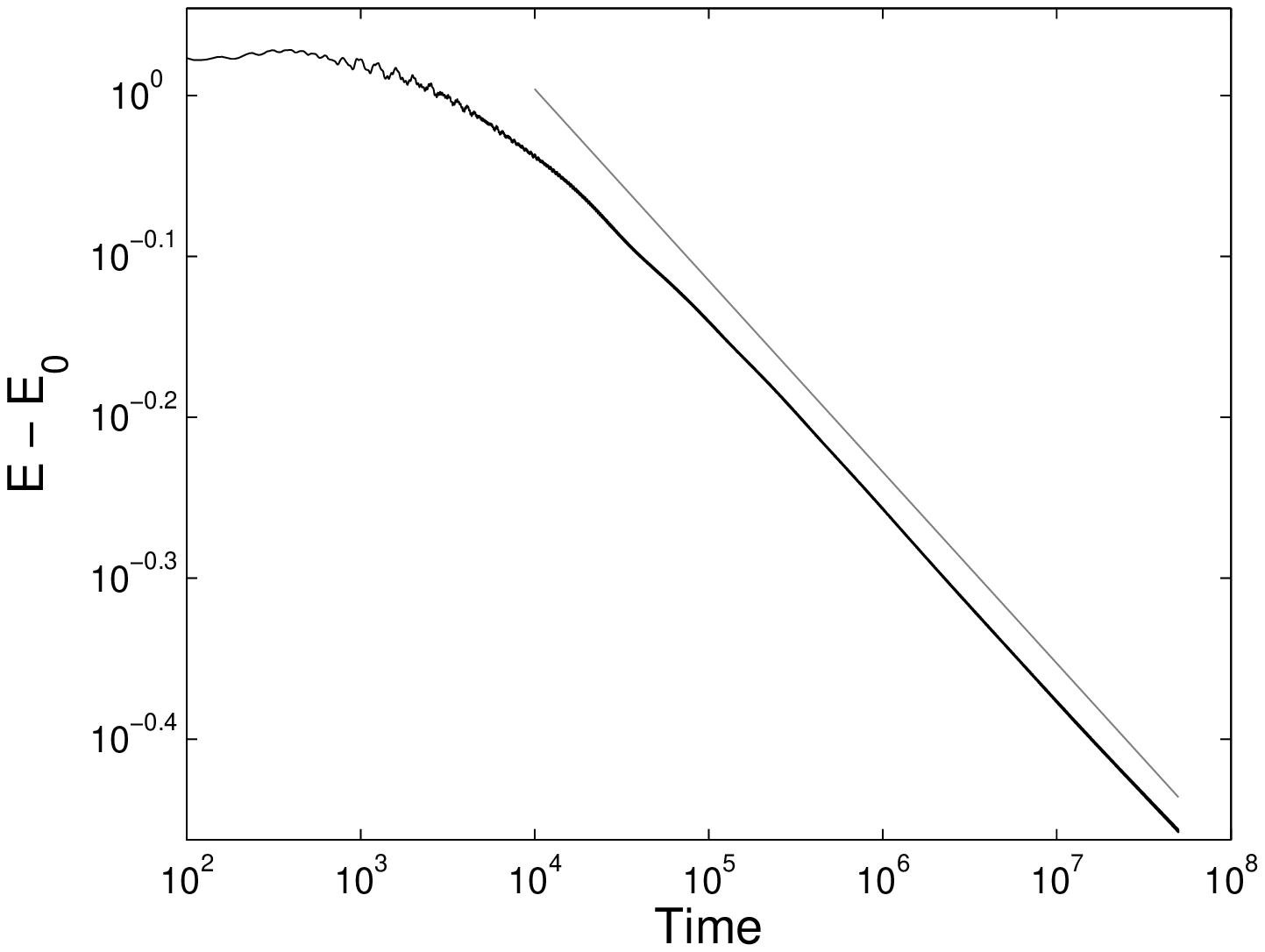}
 \caption{\label{f:sG_spherical_energy} The difference of the 
energy inside a shell of radius $r=5 r_{0}$ and the 
constant~$E_{0} = 4.769$ in sine-Gordon model in two dimensions. 
The grey straight line is a guide to eye of a power 
law $(E - E_{0}) \sim t^{-\delta}$ 
with the slope $\delta=0.119$.} 

\end{center}
\end{figure}

\begin{figure}

\centering

\includegraphics[width=0.94\hsize]{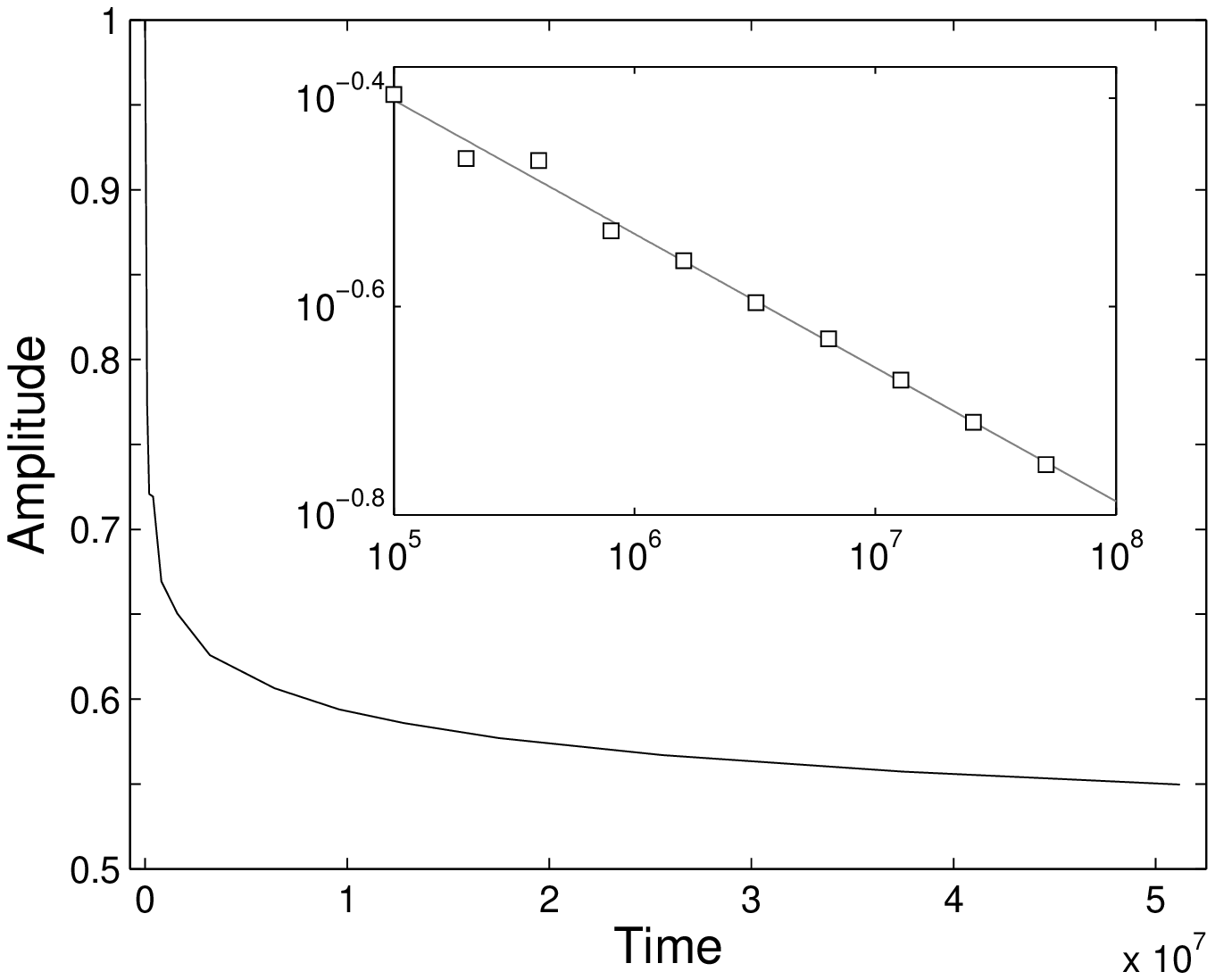}
 \caption{\label{f:sineG_amplitude} The amplitude as a 
function of time in sine-Gordon model in two dimensions. The inset shows 
the data points in the fit $A-A_0$ for $A_0=0.37$ 
for time greater than $10^5$. 
The grey straight line is the best fit power law 
$(A - A_{0}) \sim t^{-\varrho}$ with the slope $\varrho=0.13$.}
\end{figure}

\begin{figure}

\centering

\includegraphics[width=0.94\hsize]{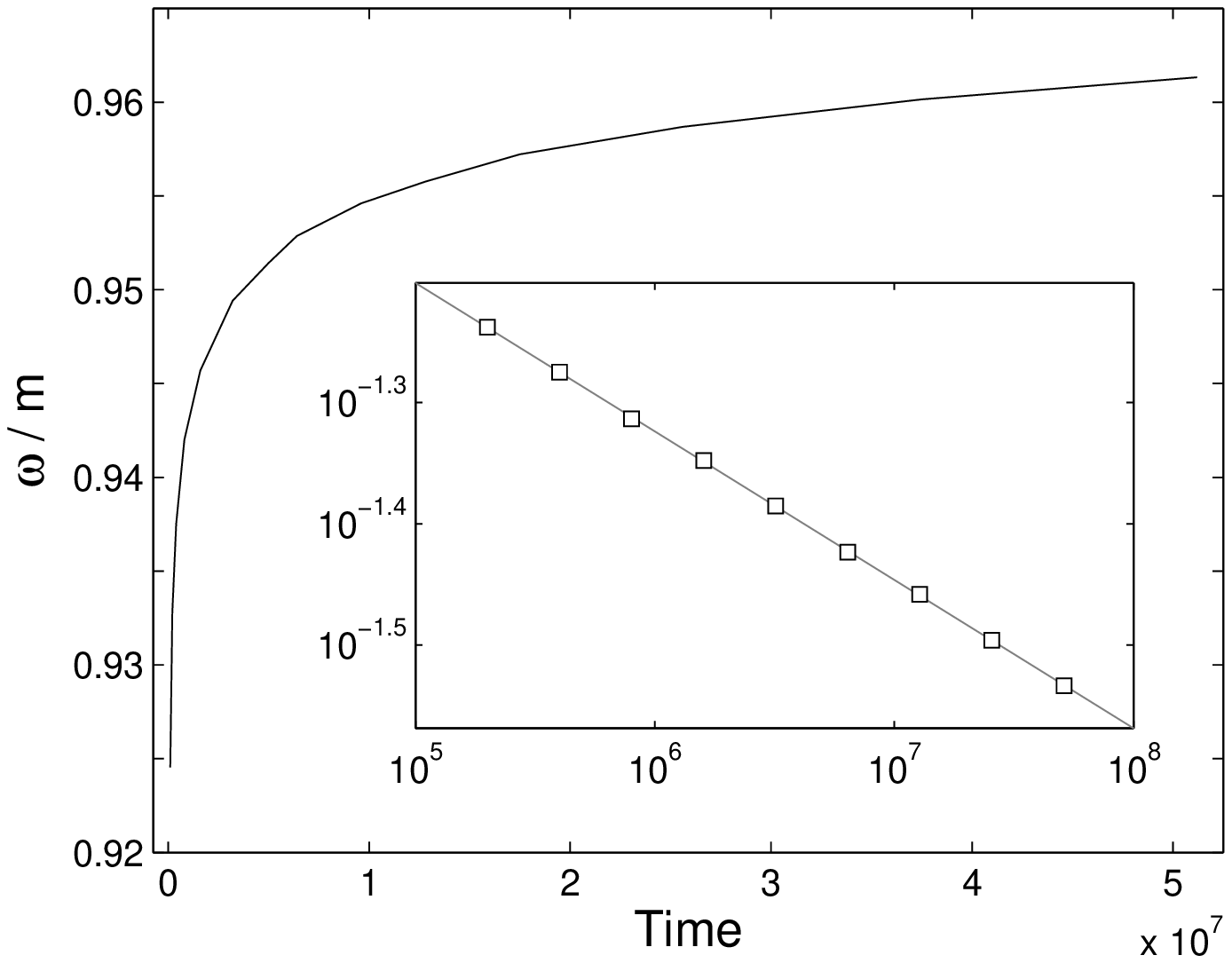}
 \caption{\label{f:sineG_frequency} The oscillation frequency as a 
function of time in sine-Gordon model in two dimensions. The inset shows 
the data points in the fit $\omega^{\star}-\omega$ for $\omega^{\star}=0.991$ 
for time greater than $10^5$. 
The grey straight line is the best fit power law 
$(\omega^{\star} - \omega_{0}) \sim t^{-\gamma}$ with the slope $\gamma=0.122$.}
\end{figure}

\begin{figure}

\centering

\includegraphics[width=0.94\hsize]{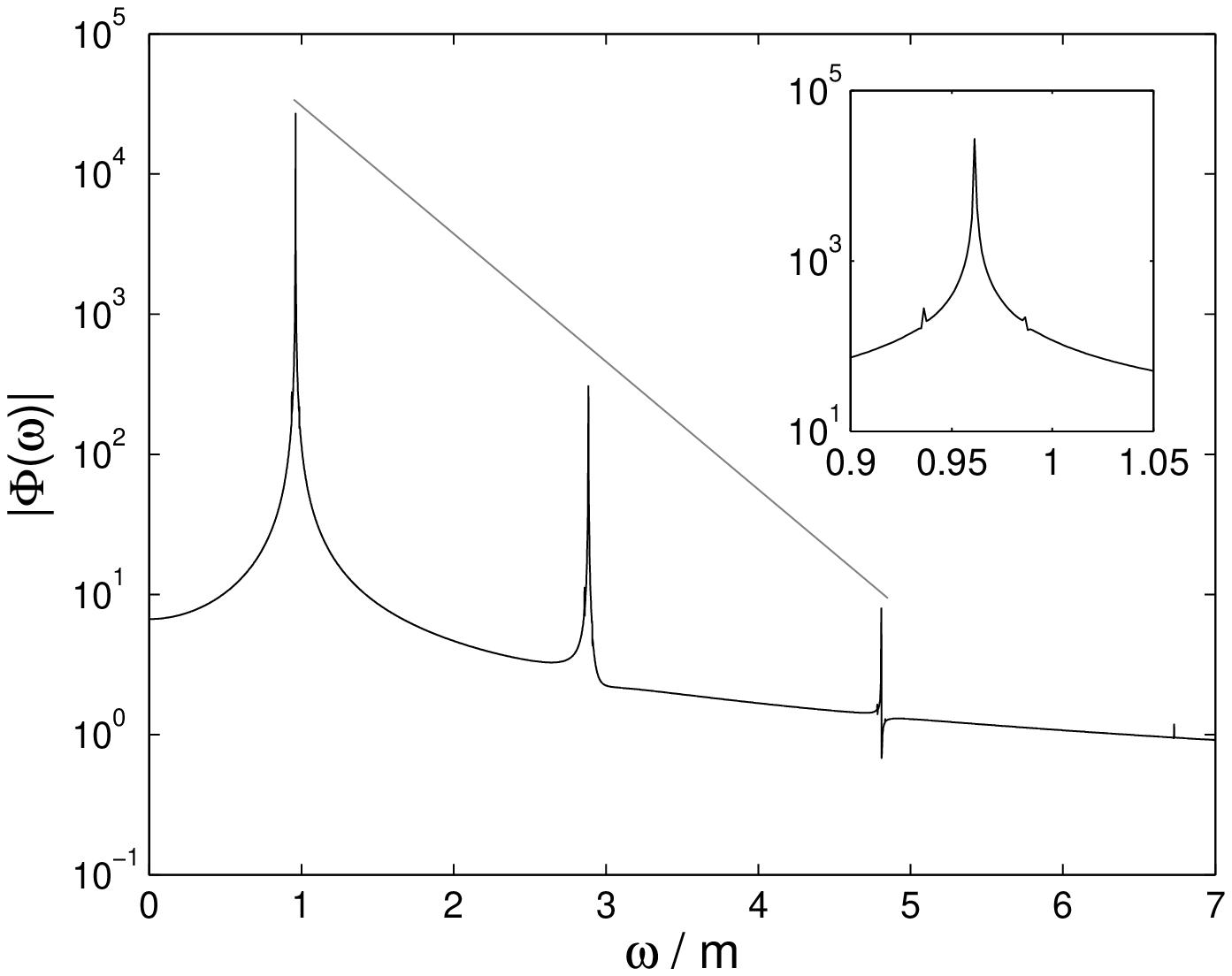}
 \caption{\label{f:sineG_spectra} The power spectrum of the field at 
the center of the oscillon, $r=0$, in sine-Gordon model in two dimensions. 
The grey straight line is a guide to eye of the exponential $\exp(-b\,\omega)$ 
with the slope $b=2.1$. The inset shows the spectrum around the oscillation 
frequency $\omega_0 \simeq 0.96$. }
\end{figure}

\begin{figure}

\centering

\includegraphics[width=0.94\hsize]{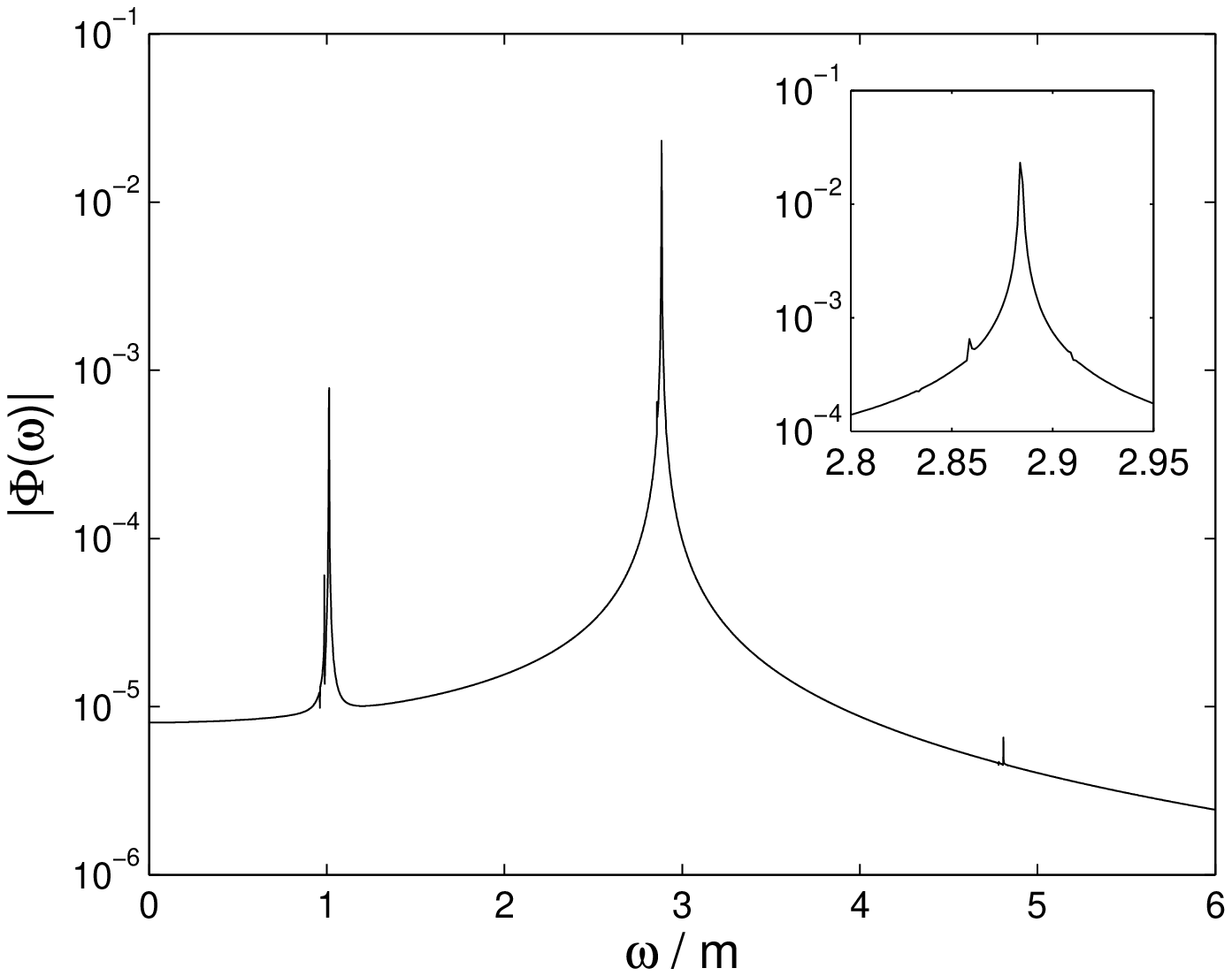}
 \caption{\label{f:sineG_spectra2} The power spectrum of the field 
far away from the oscillon core at $r=80$, in sine-Gordon model in two 
dimensions. The inset shows the spectrum around the location of the 
most prominent peak at $\omega \simeq 2.88$.}
\end{figure}

Figure~\ref{f:sG_spherical_energy0} shows the energy inside the shell 
with radius $r=5r_0$ in the simulation up to $5 \cdot 10^7$ time units. 
This is qualitatively very similar to the case in quartic theory. 
We sought again for a power law using least-square fit. The result is 
presented in Figure~\ref{f:sG_spherical_energy} where the 
constant~$E_0=4.769$ is 
subtracted from the energy inside the shell. The resulting curve has 
the slope $-0.119$ on the logarithmic scale, demonstrated by the 
grey straight line in the figure. We lack some of the dynamical range 
reached in the simulation in $\phi^4$ theory, but there is evidence 
for a power law roughly over three orders of magnitude in time 
with an exponent $\delta=0.119 \pm 0.003$. 
The rate of decrease is marginally smaller than what we observed 
in $\phi^4$ theory, and the asymptotic energy of the 
oscillon, $E_0 = 4.77 \pm 0.01$, is now slightly larger, but generally 
oscillons in both models are comparatively similar lumps of energy.

The amplitude of the oscillations behaves also very similar way 
compared with the quartic theory in two dimensions. 
Figure~\ref{f:sineG_amplitude} shows the amplitude over the course of time. 
The inset demonstrates the best-fit power law where 
$A_0 = 0.37 \pm 0.10$ and $\varrho = 0.13 \pm 0.05$. 
The derived errors in the fit are quite large, it can be seen that 
there is substantial scatter around the central value for time~$<10^6$, 
the oscillon in sine-Gordon model has a larger asymptotic 
amplitude~$A_0$ than in $\phi^4$ theory.

The evolution of the frequency as a function of time is shown 
in Figure~\ref{f:sineG_frequency}. Like in the case of energy, 
the rate of change is less than in quartic theory and the oscillation 
frequency reaches value $\omega=0.961$ at the end of the simulation 
at time point $t=5.12 \cdot 10^7$. 
The search for a critical frequency and an associated power law here 
yields quite a stringent results. There is a remarkably good fit to the data 
with $\omega^{\star} = 0.991 \pm 0.002$ and 
$\gamma=0.120 \pm 0.004$.
The data points for $\omega^{\star} - \omega(t)$ are shown in the inset 
of Figure~\ref{f:sineG_frequency} together with the grey straight line 
that demonstrates the power law~$t^{-\gamma}$. 
This suggests now that there exists a critical 
frequency $\omega^{\star} < 1$ in two-dimensional sine-Gordon model 
that would correspond an oscillon with a minimum energy.

The power spectrum of the oscillon in sine-Gordon model is shown 
in Figure~\ref{f:sineG_spectra} obtained by 
a Fourier transform in an interval of length $5000$ in time units starting 
at time $t=5.12 \cdot 10^7$.
There are no even multiples of the oscillation frequency $\omega_0$ 
present in the spectra. This is simply due to the symmetry of the 
potential~(\ref{sine-Gordon}) in sine-Gordon model with respect 
to its minima - there cannot be non-zero terms $f_{n}(r)$ 
in~(\ref{solution-ansatz}) for an even $n$. 
The remaining peaks are exponentially suppressed as 
a function of frequency~$\omega$. This suppression is even quantitatively 
very similar to the one in quartic theory. The small difference in the 
slope we measure is well within the limits of uncertainty we can expect 
in evaluating the height of the narrow peaks.

Figure~\ref{f:sineG_spectra2} shows the Fourier transform of the field 
in the same time 
interval as Figure~\ref{f:sineG_spectra}, but at location $r=80$. 
Now the most dominant peak is at $\omega=2.88=3\omega_0$, the first 
radiative mode as the even multiples of $\omega_0$, including $2\omega_0$, 
do not exist. 
There is still a visible pinnacle at $5\,\omega_0$ and a substantial 
peak just above the radiation frequency $\omega=1$, but exactly like 
in $\phi^4$ theory that one is more than order of magnitude suppressed 
compared with the prominent one at the frequency $3\omega_0$. 
Also in sine-Gordon model it is the first non-zero multiple of the basic 
frequency~$\omega_0$ above the threshold 
for radiation that leaks the energy from the oscillon.

The Gaussian ansatz~(\ref{gaussian-ansatz}) does not create long-lived 
oscillons in sine-Gordon model in three dimensions. 
For an investigation with fine-tuned initial conditions 
see~\cite{Piette:1997hf}.


\section{A Convex Potential}

\begin{figure}
\begin{center}

\includegraphics[width=0.94\hsize]{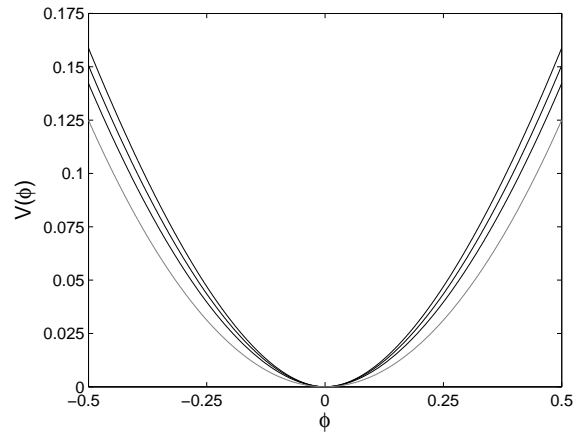}
 \caption{\label{f:potential} Potential $V(\phi)$ in~(\ref{convex_potential}) 
around the minimum for values $p=0.4,0.3,0.2$ (from top to bottom). 
They grey line shows the quadratic potential $\phi^2/2$. } 

\end{center}
\end{figure}

\begin{figure}
\begin{center}

\includegraphics[width=0.94\hsize]{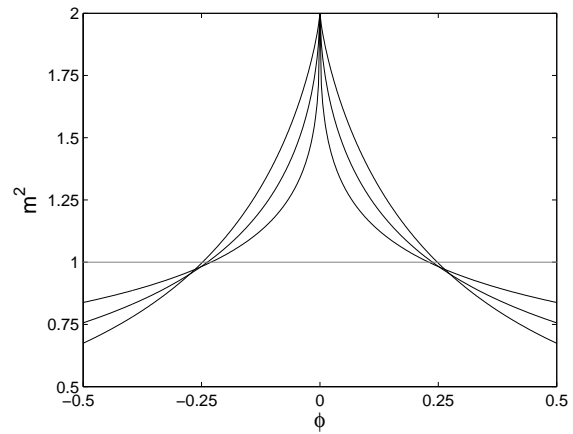}
 \caption{\label{f:mass_square} The second derivative of the 
potential $V(\phi)$ in~(\ref{convex_potential}) for $p=0.4,0.3,0.2$ 
(from broader to narrower spikes). 
The grey line shows the constant $m^2=1$ for a quadratic potential.} 

\end{center}
\end{figure}

Majority of the studies on oscillons has been carried out in the 
context of $\phi^4$ theory (see 
e.g.~\cite{Gleiser:1993pt,Copeland:1995fq,Fodor:2006zs,Saffin:2006yk}) 
or in sine-Gordon model.
Both potentials~(\ref{quartic}) and~(\ref{sine-Gordon}) have 
degenerate vacua and inflections points.
In~\cite{Kasuya:2002zs} the study dealt potentials that are nearly quadratic. 
In particular a quadratic potential with a negative logarithmic correction was 
reported to support long-lived lumps (called I-balls by the authors). 
Very recently there has been interest in convex potentials and a 
study~\cite{Amin:2011hj} considered the formation of oscillons 
in the following potential
\begin{eqnarray}
V( \phi ) = \frac{m^2 M^2}{2 \beta} 
\left[ \left( 1+ \frac{\phi^2}{M^2}\right)^{\beta} -1 \right] \, ,
\label{convex_potential_2}
\end{eqnarray}
which is motivated by a number of supergravity and superstring scenarios.

It was the work in~\cite{Kasuya:2002zs} that 
initially lead us to study the effects of the form of potential on 
the existence of long-lived oscillating solutions.
We have chosen to study the following convex function
\begin{eqnarray}
V( \phi ) = \frac{\phi^2}{1+\phi^{2p}} \, ,
\label{convex_potential}
\end{eqnarray}
which reduces to quadratic when exponent $p=0$.
Figure~\ref{f:potential} shows the potential~(\ref{convex_potential}) 
close to its minimum for values $p=0.4,0.3$ and $0.2$ together with 
the quadratic potential $\phi^2 / 2$ indicated by the grey line.
The potential is steeper than quadratic one in the vicinity of the 
minimum where oscillations of the field take place, $|\phi|<1$.

The potential~(\ref{convex_potential}) sets far greater challenge 
for the absorbing boundary conditions than $\phi^4$ theory or 
sine-Gordon model. This becomes readily apparent in 
Figure~\ref{f:mass_square} where the second derivative of the potential 
is shown for the same values of parameter $p$. 
Already small deviations from the vacuum $\phi=0$ create 
large variations in $V''$, the greater the smaller parameter $p$ is.
This is a drawback from the point of view of the absorbing boundary 
conditions used because they are based on the assumption that 
the potential can be linearised in the equation of motion. 
Once the fluctuations create large corrections,  
the absorption can be expected to be only partial or, in the worst case,
the update on the boundary even pumps energy into the lattice.
We restrict considerations from now on to values $1/2 > p \geq 0.4$ 
which yield less stark variations and for which the radiation is 
clearly adequately removed from the lattice.
Potential~(\ref{convex_potential}) is also more time consuming to evaluate 
numerically and consequently the simulations do not reach such a long final 
time as the one presented in $\phi^4$ theory, but typically order 
of $10^6$ time units.

\subsection{Properties in two dimensions}

\begin{figure}
\begin{center}

\includegraphics[width=0.94\hsize]{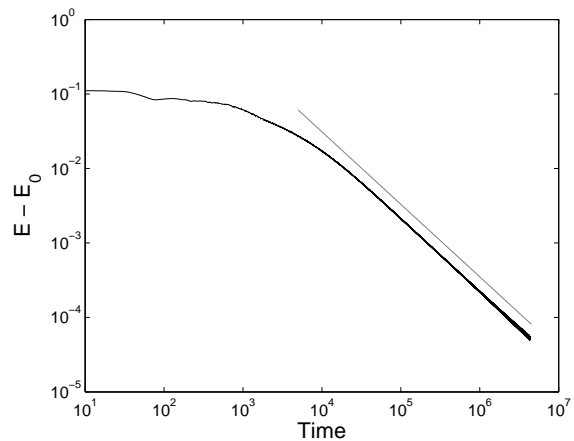}
 \caption{\label{f:subquadratic_spherical_energy1} The difference of the 
energy inside a shell of radius $r=5r_{0}$ and the 
constant~$E_{0} = 3.80343$ in the convex 
potential~(\ref{convex_potential}) for $p=0.45$ in two dimensions. 
The grey straight line is a guide to eye of a power 
law $(E - E_{0}) \sim t^{-\delta}$ with the slope $\delta = 0.97$.} 

\end{center}
\end{figure}

Also here the oscillon is initialised by a Gaussian 
deviation from the vacuum 
\begin{eqnarray}
\phi(r) =  C \cdot \exp(-r^2/r_{0}^2),
\label{gaussian-ansatz2}
\end{eqnarray}
with the same width as before $r_0 \simeq 2.9$, 
but due to the steepness of the potential~(\ref{convex_potential}) 
we set the amplitude of 
the displacement to be smaller, $C=0.6$. The oscillons formed are 
well described by a Gaussian shape though with a slightly larger width 
and a longer tail. After a short initial phase of radiation the oscillon 
settles into a state characterised by a very constant energy. 
We present the data from simulation when the exponent was set to $p=0.45$. 
The energy inside the radius $r=5r_0$ for this choice remains above $E=3.8$. 
We carried out the least-square fit for the energy. Now far more fine-tuning 
is required for seeking the asymptotic energy $E_0$.
Figure~\ref{f:subquadratic_spherical_energy1} shows the constant~$E_0=3.80343$ 
subtracted from the energy inside the shell. 
The radiative component decays now almost inversely proportional to time, 
the best fit exponent being $\delta=0.97$ and the corresponding power law 
demonstrated by the grey line in the figure. There is agreement with this 
power law over two orders of magnitude in time during which the difference 
$\left( E(t)- E_0 \right)$ drops below $10^{-4}$.
The best-fit values for the asymptotic energy and the exponent 
governing the power law are 
$E_ 0 = 3.80343(4)$ and $\delta = 0.974 \pm 0.003$, respectively.
The diminutive uncertainty in the value of energy reflects the 
fine-tuning needed in the fit, and in turn the stability of the value 
the energy reaches in the simulation.

Nearly constant energy implies also very stable oscillation frequency. 
After initial phase the oscillation frequency settles to 
$\omega_0=0.879$ and we do not observe any increase in this, but within 
the accuracy to determine frequencies it remains 
the same until the end of the simulation beyond time $t=4 \cdot 10^6$. 
It is noteworthy that this value is 
considerably below the threshold for radiation. 
The amplitude of the oscillations settles to a value $A \simeq 0.49$.

\begin{figure}
\begin{center}

\includegraphics[width=0.94\hsize]{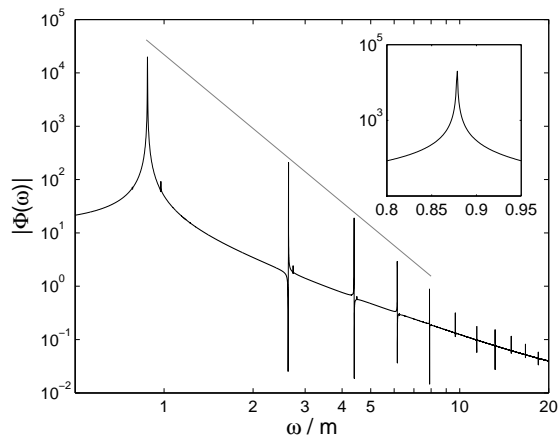}
 \caption{\label{f:subquadratic_2d_spectra}  
The power spectrum of the field at the center of the oscillon, 
$r=0$, in the convex potential~(\ref{convex_potential}) for $p=0.45$ in two 
dimensions. The grey straight line is a guide to eye of the power law 
$\omega^{-\kappa}$ with the slope $\kappa=4.6$. The inset 
shows the spectrum around the oscillation 
frequency $\omega_0 \simeq 0.88$.} 

\end{center}
\end{figure}

\begin{figure}
\begin{center}

\includegraphics[width=0.94\hsize]{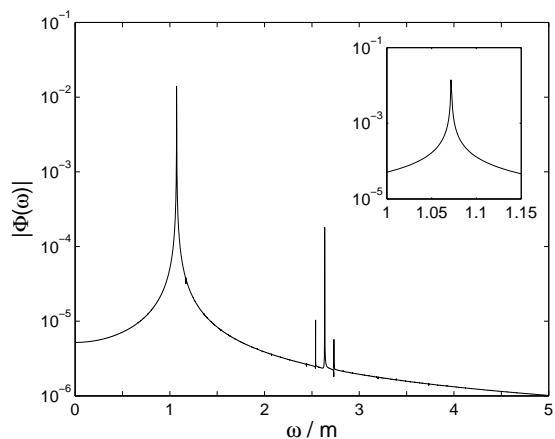}
 \caption{\label{f:subquadratic_2d_spectra2} 
The power spectrum of the field far away from the oscillon core at $r=80$, 
in the convex potential~(\ref{convex_potential}) for $p=0.45$ in two 
dimensions. The inset shows the spectrum around the most prominent peak 
at frequency $\omega \simeq 1.07$.} 

\end{center}
\end{figure}

We show the power spectrum of the oscillon in the 
potential~(\ref{convex_potential}) for $p=0.45$ obtained in the 
previously used time interval starting at time $t=4 \cdot 10^6$ 
in Figure~\ref{f:subquadratic_2d_spectra}. 
Because the potential~(\ref{convex_potential}) is symmetric around its 
minimum, there are only peaks at odd multiples of the oscillation 
frequency as in sine-Gordon model. But it is surprising that the suppression 
is governed by a power law, we plot also the frequency axis on a logarithmic 
scale in Figure~\ref{f:subquadratic_2d_spectra}. 
On the basis of the stability and weak radiation from the oscillon 
a strong exponential suppression of radiative modes would have been 
likely. The power law decrease as a function of the frequency is 
considerably quick though, the amplitudes are suppressed roughly 
as $\omega^{-5}$ though the exponent should be considered only indicative.

Figure~\ref{f:subquadratic_2d_spectra2} shows the frequencies present far 
away from the oscillon core at $r=80$. There is a peak at 
frequency~$\omega=2.64$, thus at $3\omega_0$, but this one is not the 
dominant. Instead the amplitude of the peak just above the radiation 
frequency at $\omega=1.07$ is over seventy times higher. The dominant 
radiative mode is not related to multiples of the oscillation 
frequency~$\omega_0$.

\subsection{Properties in three dimensions}

\begin{figure}
\begin{center}

\includegraphics[width=0.94\hsize]{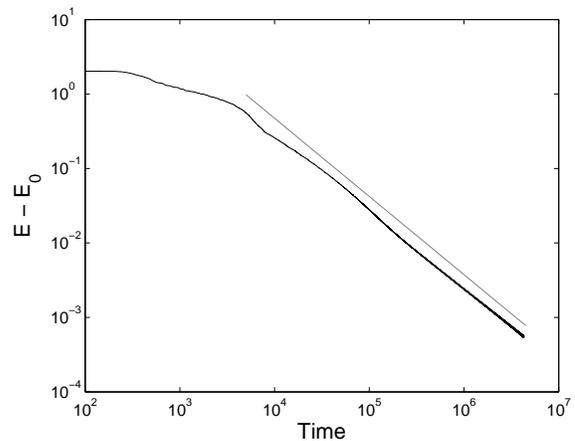}
 \caption{\label{f:subquadratic_energy_3d} The difference of the 
total energy and the constant~$E_{0} = 13.908$ in the convex 
potential~(\ref{convex_potential}) for $p=0.45$ in three dimensions. 
The grey straight line is a guide to eye of a power 
law $(E - E_{0}) \sim t^{-\delta}$ with the slope $\delta = 1.05$.} 

\end{center}
\end{figure}

We use the same Gaussian initial ansatz~(\ref{gaussian-ansatz2}) 
with the same width and amplitude as in two dimensions. 
This initial profile creates now an oscillating lump also in three 
dimensions though the oscillon has a larger width and a longer tail 
than the Gaussian distribution. 
The data shown is for the same value $p=0.45$ in the 
potential~(\ref{convex_potential}). After initial stronger radiative 
phase, there remains energy above the value $E=13.9$ within the 
radius $r=5r_0$. 

For the least-square fit we use now the data on the total energy in the 
lattice because the oscillon is more spatially spread and due to the 
precision required in determining the value of the constant $E_0$ 
below $10^{-3}$ the fluctuations in the energy inside the shell become 
apparent (they are to a lesser extent already visible 
in Figure~\ref{f:subquadratic_spherical_energy1}).
Figure~\ref{f:subquadratic_energy_3d} shows the constant $E_{0}=13.908$ 
subtracted from the total energy. The decaying component decreases 
fast in time, 
the best-fit for the exponent governing the power law 
is $\delta=1.09 \pm 0.01$ while the asymptotic value of the 
energy is $E_0 = 13.9081 \pm 0.0002$. There is fairly large uncertainty 
in the value of the exponent, but there is good evidence that the radiative 
part decays faster than inversely proportional to time. 
The amplitude related to this energy is $A \simeq 0.34$.

\begin{figure}
\begin{center}

\includegraphics[width=0.94\hsize]{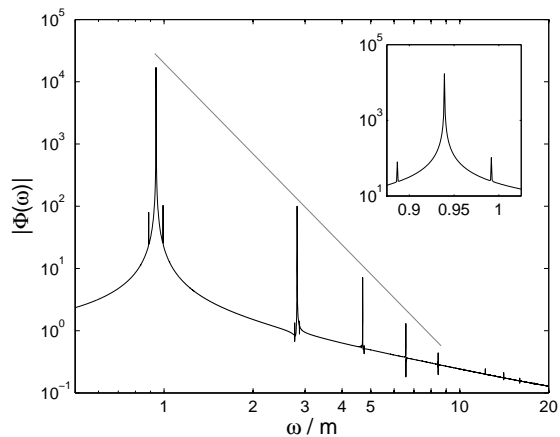}
 \caption{\label{f:subquadratic_3d_spectra} 
The power spectrum of the field at the center of the oscillon, 
$r=0$, in the potential~(\ref{convex_potential}) for $p=0.45$ in three 
dimensions. The grey straight line is a guide to eye of the power law 
$\omega^{-\kappa}$ with the slope $\kappa=4.85$. The inset 
shows the spectrum around the oscillation 
frequency $\omega_0 \simeq 0.94$.} 

\end{center}
\end{figure}

\begin{figure}
\begin{center}

\includegraphics[width=0.94\hsize]{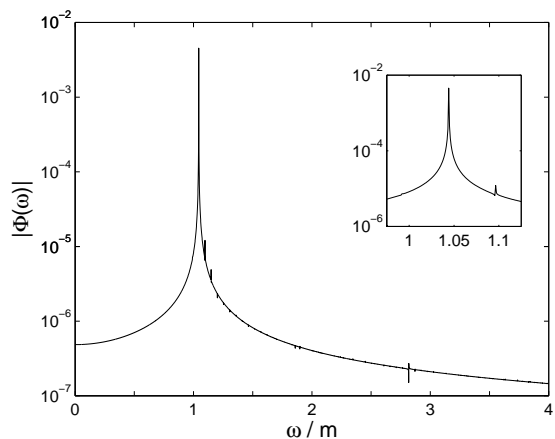}
 \caption{\label{f:subquadratic_3d_spectra2} 
The power spectrum of the field far away from the oscillon core at $r=80$, 
in the potential~(\ref{convex_potential}) for $p=0.45$ in three 
dimensions. The inset shows the spectrum around the most prominent peak 
at frequency $\omega \simeq 1.04$.} 

\end{center}
\end{figure}

With the aforementioned energy the oscillation frequency 
settles to $\omega \simeq 0.939$. 
Again we do not observe increase in the frequency within the numerical 
accuracy after the initial phase.
Figure~\ref{f:subquadratic_3d_spectra} shows the power spectrum of the 
three-dimensional oscillon obtained in the previously used time interval 
starting at time $t= 4 \cdot 10^6$. The result is very similar to 
Figure~\ref{f:subquadratic_2d_spectra} with peaks at odd multiples of 
the basic frequency $\omega_0$. The suppression of the higher amplitudes 
is again governed by a power law demonstrated by the grey straight line 
in the figure. The measured slope is slightly steeper 
than in two dimensions, but this difference is hardly significant taking 
into account the accuracy of the data we can expect.

Figure~\ref{f:subquadratic_3d_spectra2} shows the frequencies at the 
location $r=80$. 
There is only barely visible structure above the background 
at $\omega \approx 2.82$, three times the value of the oscillation frequency. 
The only peak of any note is located at frequency $\omega=1.04$, thus barely 
above the threshold for propagating radiation.


To summarize our findings of the oscillons in the convex 
potential~(\ref{convex_potential}), the change from two to three dimensions 
seem to have surprisingly small effect. In both dimensions there remains a 
very stable lump of energy oscillating at a constant frequency which emits 
only tiny amount of radiation. 
While we do not have an explanation for their longevity, the investigation in 
frequency space gives one answer.
As the oscillon can primarily only excite modes that are just above 
the threshold 
for radiation, i.e. $\left(\omega -m\right) \ll 1$, obviously such modes 
can carry only lesser amount of energy away from the core of the lump. 
Consequently we can expect far reduced emitted radiation compared with 
$\phi^4$ theory and sine-Gordon model as we indeed observe (it cannot 
be excluded that some of the signal we see is from relic radiation 
on the lattice that was not absorbed by the boundary, but reflected back 
to the grid).

We have considered here only very limited range of values of the parameter~$p$ 
in the potential~(\ref{convex_potential}), but the potential supports 
oscillons in a wider variety of the values of the exponent $p$ governing 
its steepness~\cite{Robert}. With a different choice for lattice spacing 
and time step, oscillons have survived up to $10^7$ time units~\cite{Robert}. 
This is hardly a surprising result taking into account their 
minuscule radiation losses.

Finally, it should be noted that non-radiating oscillons has been found in 
one-dimensional signum-Gordon model~\cite{Arodz:2007jh} 
(more recently also a new class of swaying oscillons have 
been discovered in the same model~\cite{Arodz:2011zm}). 
It is defined by the potential
\begin{eqnarray}
V(\phi) = g \, |\phi|.
\label{signum-Gordon}
\end{eqnarray}
The convex potential~(\ref{convex_potential}) reduces to linear for $p=1/2$ 
in the limit of large field, $|\phi| \gg 1$. This is not the realm where 
oscillations of the field take place, but it cannot be excluded right away 
that there were no connection between oscillons we have 
reported here and those described in~\cite{Arodz:2007jh}. 
At very least, both are examples of oscillons appearing in theories where 
the potential is not a smooth one.

\section{Conclusions}

\begin{table}[tbp]
 \begin{tabular}{|c|c|c|c|}
   \hline
   model & dim & $E_0$ & $\delta$ \\ \hline
   quartic theory & $2$ & $ 4.613 \pm 0.003 $ & $  0.1520 \pm 0.0005 $  \\ \hline
   quartic theory & $3$ & $ 41.56 \pm 0.05 $ & $ 0.65 \pm 0.01 $  \\ \hline
   sine-Gordon & $2$ & $4.77 \pm 0.01$ & $0.119 \pm 0.003$ \\ \hline
   convex potential & $2$ & $3.803430 \pm 4 \cdot 10^{-6}$ & $ 0.974 \pm 0.003 $ \\ \hline
   convex potential & $3$ & $ 13.9081 \pm 0.0002$ & $ 1.09 \pm 0.01 $ \\ \hline
 \end{tabular}
 \caption{\label{t:energy} The values of the asymptotic energy $E_0$ and the 
exponent $\delta$ in $\left(E(t) - E_0\right) \sim t^{-\delta}$ for the oscillons in the various models studied with the initial 
conditions as mentioned in the text.}
\end{table}

We have studied oscillons in $\phi^4$ theory, sine-Gordon model 
and in convex potentials, performing (1+1)-dimensional simulations 
assuming radial symmetry.
An important ingredient was the development of 
absorbing boundary conditions for a massive, real scalar 
field. Generally absorbing boundary conditions work well and 
radiation is removed from the lattice.

We showed that the time evolution of the energy and the frequency of 
oscillons is well modelled by dividing it to a constant and a decaying 
radiative part. There is strong evidence that the decay of the radiative 
part is governed by a power law in all models considered. 
We further determined the values of the constant components as well as 
values of the exponents in the power law decay, which we summarize 
in Table~\ref{t:energy}.  
The power law decay has been previously reported in $\phi^4$ 
theory~\cite{Gleiser:2008ty,Gleiser:2009ys} and for flat-top 
oscillons in~\cite{Amin:2010jq}. 
Note that for the oscillons in the convex potentials the frequency 
remains constant, which is a qualitatively new behaviour for an oscillon.

\begin{table}[tbp]
 \begin{tabular}{|c|c|c|c|}
   \hline
   model & dimension & $A_0$ & $\varrho$ \\ \hline
     quartic theory & 2 & $ 0.33 \pm 0.06 $ & $  0.13 \pm 0.02 $  \\ \hline
     quartic theory & 3 & $ -1.4 \pm 0.3 $ & $ 0.066 \pm 0.007 $  \\ \hline
     sine-Gordon & 2 & $0.37 \pm 0.10$ & $0.13 \pm 0.05$ \\ \hline
 \end{tabular}
 \caption{\label{t:amplitude} The values of the asymptotic amplitude $A_0$ 
and the exponent $\varrho$ in $\left(A(t) - A_0\right) \sim t^{-\varrho}$ for the oscillons in the various models studied with the initial 
conditions as mentioned in the text.}
\end{table}

\begin{table}[tbp]
 \begin{tabular}{|c|c|c|}
   \hline
      model & $\omega^{\star}$ & $\gamma$ \\ \hline
     quartic theory & $ 0.981 \pm 0.003 $ & $ 0.18 \pm 0.02$  \\ \hline
     sine-Gordon model & $0.991 \pm 0.002$ & $0.120 \pm 0.004$ \\ \hline
   \end{tabular}
\caption{\label{t:frequency} The values of the critical 
frequency $\omega^{\star}$ and the exponent $\gamma$ in $\left(\omega^{\star} - \omega(t)\right) \sim t^{-\gamma}$ in two dimensions for the oscillons with 
the initial conditions as mentioned in the text.}
 \end{table}

We also studied the evolution of the amplitude and the oscillation 
frequency over time, our results are summarized in 
Tables~\ref{t:amplitude} and~\ref{t:frequency}. 
The oscillon amplitude also exhibits a power law decay, 
which in three dimensions is strongly modulated by a beat frequency, 
while in the case of the convex potential it is constant.

Furthermore, we examined the power spectrum of field at the centre 
of oscillon and near the boundary.
While in $\phi^4$ theory and in sine-Gordon model there is an exponential 
attenuation of higher multiples of the oscillation frequency, 
the oscillons in the convex potential show only a power law suppression.

Understanding the manner oscillons radiate and shed their energy 
away is crucial in gaining insight for their persistence and longevity.
We studied the outgoing radiation in frequency space. The emitted 
radiation in $\phi^4$ theory and in sine-Gordon model 
is strongly peaked at the first non-zero multiple of the basic frequency, 
while in the convex potential the oscillons radiate at a low frequency 
just above the threshold to release radiation. 
Our results in quartic theory, in particular in two dimensions, 
agree well with the study of quasi breathers in the same 
model~\cite{Saffin:2006yk}. There it was identified that the emissions 
take place at the frequencies $n\omega_0$, with $n\geq2$ with the 
frequency $2\omega_0$ being the dominant one.
The quasi breather approach, i.e. ansatz~(\ref{solution-ansatz}), does 
not assign any width for the frequencies. This was considered the starting 
point in the work~\cite{Gleiser:2009ys} when dealing with the basic 
oscillation frequency. Our results indicate that there is a contribution 
from both sources, from higher modes as well as from the spread related 
to the basic oscillation frequency and once again highlight 
the richness and complexity of oscillons. An approach that would account 
for both higher frequencies present in the spectra as well as the width 
of the frequencies might capture the problem of oscillon radiation 
in full.

What do the results tell about the lifetime of oscillons? 
The appearance of the constant $E_{0}$ now raises the 
question of its physical interpretation. While it would be tempting to 
define it as the energy of a non-radiating oscillon, this 
kind of conclusion cannot be made straightforwardly.
It is namely not obvious at all that there are oscillon solutions 
at such energy. 
While the energy decreases the oscillon oscillates faster and the 
oscillation frequency $\omega_{0}$ increases. 
Once close enough to the threshold for radiation $\omega = m$ oscillon 
demises when it reaches its minimum energy state~\cite{Watkins,Saffin:2006yk}.
This was well demonstrated in $\phi^4$ theory in three dimensions, 
there the asymptotic energy $E_{0}$ is well below the energy level that 
caused the oscillon to dissolve.
Because the radiative part shows a slow decay in time in two dimensions, 
the energy oscillons have reached at the end of simulations is still well 
above the asymptotic value~$E_0$ in $\phi^4$ theory and sine-Gordon model, 
more precisely by $13\%$ and $7\%$, respectively 
(it should be noted that there are some theoretical bounds on the 
minimum energy of lumps~\cite{Buniy:2003xq}).

However, the case for the convex potentials presented here is 
definitely intriguing. We have shown that there the suppression 
of the radiation losses is quick, roughly inversely proportional to time. 
Thus only very small further decrease in energy is to be expected to 
occur, orders of magnitude less than in $\phi^4$ theory or in sine-Gordon 
model. Furthermore, the frequency of oscillons is well below the threshold 
for radiation.
This all makes it possible that these oscillons could go on indefinitely long.
Another open question remaining is if it is possible to tailor potentials 
in which the emission of radiation from oscillon is strongly suppressed 
also in higher dimensions.
The convex potential we considered is not phenomenologically particularly 
well motivated, but if this sort of oscillons can also appear in more realistic 
theories their long life-time would almost certainly make them viable to 
affect physics in the system there.


\begin{acknowledgments}

P.~S. thanks Nick Manton, Fuminobu Takahashi and 
Anders Tranberg for discussions and is grateful to Melvin Varughese 
for comments on regression analysis. 
The authors acknowledge Robert Bennett for sharing his results with them.
P.~S. is supported by the Claude Leon Foundation.

\end{acknowledgments}

\appendix

\section{Absorbing Boundaries}

In this appendix the absorbing boundary conditions are presented.
As the starting point consider the following linear wave equation
\begin{eqnarray}
\ddot{\varphi} - \nabla^2 \varphi +  \varphi \cdot V''|_{\varphi=0}  =0 \, ,
\label{appendix1}
\end{eqnarray}
where $\varphi$ is now a small perturbation around a chosen vacuum 
and $V''$ evaluated at that point. 
For example in $\phi^4$ theory $\phi \rightarrow  v - \varphi$ 
and $V''(\phi)|_{\phi=v} = 2\lambda v^2 = m^2$.
The treatment here follows closely the one presented in~\cite{Battye} for a 
massless scalar field (for absorbing boundary conditions in general 
see also~\cite{bamberger:323}).

\subsection{Cylindrical Wave}

When one assumes radial symmetry in two space dimensions the 
wave equation~(\ref{appendix1}) reads
\begin{eqnarray}
\frac{{\partial ^2 \varphi}}{{\partial t^2}} -
\frac{{\partial ^2 \varphi}}{{\partial r^2}} -
\frac{1}{r}\frac{{\partial  \varphi}}{{\partial r}} + m^2 \varphi = 0.
\label{appendix9}
\end{eqnarray}
The solutions of~(\ref{appendix9}) are Bessel functions, but considering 
an absorbing boundary at far distance ($r \gg 1$) the approximative 
solution reads
\begin{eqnarray}
\varphi
(t,r) 
= \frac{1}{\sqrt{r}} 
\exp \Big[\,i\,(\sqrt{\omega^2-m^2}\,r+\omega t \,)\Big]\,,
\label{appendix10}
\end{eqnarray}
where $\omega$ is a dual variable to $t$. 
This solution~(\ref{appendix10}) is annihilated at the boundary 
$r=r_{\rm b}$ (to order $r^{-5/2}$) by the operator in the parenthesis 
\begin{eqnarray}
\Bigg[ \frac{{\partial }}{{\partial r}} -
i \sqrt{\omega^2-m^2} +\frac{1}{2r}\,  \Bigg] \varphi_{\Big|_{r=r_{\rm b}}} = 0.
\label{appendix11}
\end{eqnarray}
From the point of view of numerical implementation using leap 
frog update we aim to second order equations at the boundary. 
Expanding the square root to second order in~(\ref{appendix11}) 
and replacing the variable $\omega$ by the corresponding 
time derivative one obtains an absorbing boundary condition 
on the edge of the one-dimensional grid
\begin{eqnarray}
\frac{{\partial }}{{\partial t}} \frac{{\partial \varphi}}{{\partial r}}
+ \frac{{\partial ^2 \varphi}}{{\partial t^2}}
- \frac{1}{2} m^2 \varphi + \frac{1}{2r} \frac{{\partial \varphi }}{{\partial t}} = 0.
\label{appendix12}
\end{eqnarray}
The equation~(\ref{appendix12}) permits a straightforward 
time evolution using leap frog algorithm.
To be more precise, the condition~(\ref{appendix12}) yields 
a new evolution equation for the field 
momentum $\Pi$ compared to that in the interior of the lattice, 
whereas the update of the field $\phi$ stays unaltered.
The last term in~(\ref{appendix12}), suppressed by inverse of physical distance~$r$, 
results from the treatment in the polar coordinates. 
It turns out to be fairly insignificant with the numerical set-up used 
in this study, but in a smaller grids it increases the absorption.

\subsection{Spherical Wave in Three Dimensions}

When one assumes spherical symmetry in three dimensions 
the solution for the wave equation 
\begin{eqnarray}
\frac{{\partial ^2 \varphi}}{{\partial t^2}} -
\frac{{\partial ^2 \varphi}}{{\partial r^2}} -
\frac{2}{r}\frac{{\partial  \varphi}}{{\partial r}} + m^2 \varphi = 0.
\label{appendix4-13}
\end{eqnarray}
is given by 
\begin{eqnarray}
\varphi
\left(t,r \right) 
= \frac{1}{r} 
\exp \Big[\,i\,(\sqrt{\omega^2-m^2}\,r+\omega t \,)\Big]\,.
\label{appendix4-14}
\end{eqnarray}
This is annihilated at the boundary 
$r=r_{\rm b}$ by 
\begin{eqnarray}
\Bigg[ \frac{{\partial }}{{\partial r}} -
i \sqrt{\omega^2-m^2} +\frac{1}{r}\,  \Bigg] 
\varphi_{\Big|_{r=r_{\rm b}}} = 0 \, ,
\label{appendix4-15}
\end{eqnarray}
which leads to following absorbing boundary condition 
\begin{eqnarray}
\frac{{\partial }}{{\partial t}} \frac{{\partial \varphi}}{{\partial r}}
+ \frac{{\partial ^2 \varphi}}{{\partial t^2}}
- \frac{1}{2} m^2 \varphi + \frac{1}{r} \, \frac{{\partial \varphi}}{{\partial t}}= 0,
\label{appendix4-16}
\end{eqnarray}
appropriate when a three-dimensional system is considered 
in a spherically symmetric geometry.


\bibliography{oscillon}
\include{oscillon}

\end{document}